\documentclass[letterpaper,fleqn,usenatbib]{mnras}

\usepackage[T1]{fontenc}
\usepackage{ae,aecompl}

\usepackage{graphicx}	
\usepackage{amsmath}	
\usepackage{amssymb}	

\usepackage{setspace}

\title[RT Unstable Flames]{Rayleigh-Taylor Unstable Flames at Higher Reynolds Number}

\author[E. P. Hicks]{
E. P. Hicks,$^{1}$\thanks{E-mail: eph2001@columbia.edu}
\\
$^{1}$Epsilon Delta Labs, Evanston, IL, 60201, USA
}

\date{Accepted 2019 July 18. Received 2019 July 13; in original form 2018 November 7.}

\pubyear{2019}

\begin{document}
\label{firstpage}
\pagerange{\pageref{firstpage}--\pageref{lastpage}}
\maketitle
\raggedbottom
\hbadness=99999
\vfuzz=20pt


\begin{abstract}
Rayleigh-Taylor (RT) unstable flames are a key component of Type Ia and Iax supernovae explosions, but their complex hydrodynamics is still not well understood.  These flames are affected not only by the RT instability, but also by the turbulence it generates.  Both processes can increase the flame speed by stretching and wrinkling the flame.  This makes it hard to choose a subgrid model for the flame speed in full star Type Ia or Iax simulations.  Commonly used subgrid models get around this difficulty by assuming that either the RT instability or turbulence is dominant and sets the flame speed.   In previous work, we evaluated the physical assumptions and predictive abilities of these two types of models by analyzing a large parameter study of 3D direct numerical simulations of RT unstable flames.  Surprisingly, we found that the flame dynamics is dominated by the RT instability and that RT unstable flames are very different from turbulent flames.  In particular, RT unstable flames are thinner rather than thicker when turbulence is strong.  In addition, none of the turbulent flame speed models adequately predicted the flame speed.  We also showed that the RT flame speed model failed when the RT instability was strong, suggesting that geometrical burning effects also influence the flame speed.  However, these results depended on simulations with $Re\lesssim720$.  In this paper, we extend the parameter study to higher Reynolds number and show that the basic conclusions of our previous study still hold when the RT-generated turbulence is stronger. 

\end{abstract}


\section{Introduction}

Thermonuclear burning begins in single-degenerate scenarios for Type Iax and Type Ia supernovae as a very thin reacting front that propagates subsonically.  The details of this deflagration (also called a ``flame'') determine the abundance of elements produced in the explosion and the overall explosion energy.  In the leading scenario for Type Iax supernovae, a pure deflagration explosion is triggered by helium accretion onto a C/O or C/O/Ne white dwarf~\citep{branch2004, jha2006, phillips2007, jordan2012, kromer2013, fink2014, long2014, magee2016, jha2017}.  The deflagration is the only nuclear burning that takes place and it determines observables like luminosity, ejecta velocities, and how well-mixed the fusion products are.  

If the deflagration is followed by a self-sustaining, supersonic burning wave (a detonation), a normal Type Ia supernova may occur~\citep{blinnikov1986, woosley1990,khokhlov1991, gamezo2003, gamezo2004, ropke2007}.  In this case, the deflagration is still important, not only because it burns some of the star's fuel, but also because it determines how much the star expands before the detonation takes place.  Both of these factors influence the ultimate nuclear yield and energy of the explosion~\citep{seitenzahl2013}.  

The deflagration may also determine how the detonation ultimately takes place.  It is still unknown whether the deflagration can somehow transition directly to a detonation (a deflagration-to-detonation transition, or DDT), or whether the detonation is triggered in some other way.  In one commonly evoked scenario, a DDT occurs by the Zel'dovich gradient mechanism~\citep{zeldovich1970} after the flame is thickened by turbulence~\citep{khokhlov1997,khokhlov1997a, niemeyer1997,lisewski2000,dursi2006,woosley2007,woosley2009,seitenzahl2009}.  Another possibility is that the flame undergoes some type of turbulent self-acceleration~\citep{poludnenko2011,poludnenko2015,poludnenko2017_icders}.  In both of these cases, how the detonation takes place depends on the conditions produced by, and the nature of, the deflagration.  Ultimately, determining single-degenerate Type Iax and Ia nuclear yields, luminosity and ejecta properties and evaluating possible detonation mechanisms for Type Ia supernovae require a full understanding of supernovae flames.

However, understanding deflagrations in Type Ia and Iax supernovae is difficult because they are hydrodynamically complex.  In scenarios where the deflagration begins near the center of the star and burns outward it is Rayleigh-Taylor (RT) unstable \citep{rayleigh1883,taylor1950}  because the dense fuel that the flame consumes rests above the lighter ashes that it produces.    The RT instability stretches and wrinkles the surface of the flame, increasing its surface area and speeding it up.  This deformation of the flame surface also baroclinically generates turbulence, which back-reacts on the flame front \citep{V05,zhang2007,hicks2013, hicks2015}.  The turbulence may further stretch and wrinkle the flame and may also affect the flame structure.  A secondary Kelvin-Helmholtz instability can also wrinkle the flame.   In addition, the flame may be in a complex environment.  If the flame is initiated in the convective core of the star, it will propagate through convective turbulence~\citep{nonaka2012}.  If the initial spark is large enough, burning could take place on the surface of a growing buoyant bubble and the flow of fluid around the bubble could modify the RT instability~\citep{vladimirova2007,zingale2007,aspden2011a}.  Alternatively, ignition could be geometrically complex and occur in multiple sparks~\citep{seitenzahl2013, fink2014, long2014}.  Finally, magnetic fields will also affect the flame dynamics if they are strong enough~\citep{hristov2018}. 

All of these complications make it hard to choose a subgrid model for the flame speed in full star simulations of Type Ia and Iax supernovae.   Subgrid models are necessary because the size of the white dwarf (Earth-sized) is so much larger than the typical flame width ($10^{-4}$ to $10^{2}$ cm, \citet{timmes1992}) that both cannot be resolved in the same simulation.  Two basic types of subgrid models have been used.  RT-type subgrid scale (RT-SGS) models \citep{khokhlov1995,khokhlov1996,gamezo2003,gamezo2004,gamezo2005, zhang2007,townsley2007,jordan2008} are based on the assumption that RT stretching and wrinkling of the flame front determines the flame speed.  The flame front is assumed to be self-similar (fractal) so that the velocity at any unresolved scale, $\Delta$, is given by the velocity $v_{RT}(\Delta) \propto \sqrt{g \, A \, \Delta}$  naturally associated with the Rayleigh-Taylor instability at the length scale $\ell = \Delta$.  Here, $g$ is the gravitational acceleration and the Atwood number is $A = (\rho_{\rm fuel}-\rho_{\rm ash})/(\rho_{\rm fuel}+\rho_{\rm ash})$, where $\rho_{\rm fuel}$ and $\rho_{\rm ash}$ are the densities of the fuel and the ash. There is a competition between the creation of surface area by the RT instability and destruction of surface area by burning in highly curved or densely packed parts of the flame surface.  This ``self-regulation'' forces the flame to propagate at the RT flame speed on average~\citep{khokhlov1995, zhang2007}.  

The second type of model, turbulence-based subgrid scale (Turb-SGS) models, is based on the assumption that interaction between turbulence and the flame front sets the flame speed and determines the flame's behavior~\citep{niemeyer1995,niemeyer1997,niemeyer1997a,reinecke1999,ropke2005,schmidt2006a,schmidt2006b,jackson2014}.  Models of this type are adapted from premixed turbulent combustion theory, which is the study of flames propagating through pre-existing turbulence.  In these models, the flame speed is typically some function of the rms velocity of the upstream turbulence.  Turb-SGS models implicitly assume that the flame interacts with its own self-generated turbulence in the same way as it would interact with upstream turbulence.  Both types of models are educated guesses  about how fast the flames should propagate, but they are based on different physical reasoning. The only way to determine whether either model type makes good predictions is to test them.

There have been many studies of RT unstable flames~\citep{khokhlov1994,khokhlov1995, bell2004b,V03,V05,zingale2005a,zhang2007,ciaraldi-schoolmann2009,chertkov2009,biferale2011,hicks2013,hicks2014,hicks2015,hristov2018}, but the only parameter study that was large enough to test the predicted flame speed scalings and also resolved the viscous scale was our study,~\citet{hicks2015}.   This parameter study included simulations at 11 different parameter combinations, 6 with a smaller non-dimensional domain width ($L=32$) and 5 with a larger domain width ($L=64$).  We varied the non-dimensional gravity ($G$) so that the flame ranged from a simple rising bubble burning at a constant speed to a complex, highly wrinkled surface with a highly variable flame speed.  The Reynolds numbers calculated for these simulations ranged from $70$ to $720$.  To isolate the effect of the RT instability on the flame front, we made as many simplifications as possible including using a simple model reaction instead of the full reaction chain and using the Boussinesq approximation to ignore compressibility effects.  Finally, we focused on the saturated state that the flame reaches when it is confined by the sides of the simulation domain.  In this state, the flame speed (and other quantities) oscillate around a statistically steady average as the flame self-regulates.   We used long time series to extract robust scalings that did not depend on time and could be compared with the flame speed models.

The results of this parameter study were surprising and suggestive.  First, we measured the flame width to determine whether RT unstable flames follow the regime predictions of turbulent combustion theory.  When turbulence is weak, the flame should be in the flamelets regime and have the same width as a laminar flame.  When turbulence is strong, the flame should transition to the thin reaction zones regime and have a width greater than the laminar flame width.  Instead, we found that the flames became thinner as the turbulence became stronger!   RT stretching overcomes any diffusive thickening by small turbulent eddies.   This result cast doubt on the wisdom of using the predictions of turbulent combustion theory to formulate subgrid models for RT unstable flames.  In addition, it makes theories for the DDT in Type Ia that rely on the transition to thin reaction zones, like the Zel'dovich gradient mechanism, seem unlikely.  Second, we measured the flame speed and the turbulent rms velocity and found an unusual relationship between the two.  Specifically, the flame speed grows faster than linearly with the rms velocity, so the curve on the burning velocity diagram (the plot of $s$ vs $u'$) is concave up.   However, models for the flame speed in traditional turbulent combustion theory are typically either linear or concave down.  The unusual shape of our measured data meant that all three types of turbulent combustion models that have been adapted for Type Ia simulations (linear, scale invariant, and bending) failed to fit the data well.  Overall, our results implied that RT unstable flames do not behave like turbulent flames because the RT instability dominates the flame dynamics.   This should have boded well for the RT-SGS model, but, although that model fit the data well when the RT instability was weak, it underpredicted the flame speed when the RT instability was strong.  This led us to hypothesize that some geometric factor, like enhanced local burning in regions of high curvature (cusps), increased the global flame speed above the RT prediction.  Ultimately, neither type of model consistently predicted the flame speed.    

However, all of our unusual findings in~\citet{hicks2015} were based on data collected at $Re\lesssim720$.  Do these results still hold when the RT-generated turbulence is stronger?  In order to address that question, we added two new simulations with $L=64$, $G=16$ that had measured average Reynolds numbers of $Re=966$ and $985$ to the parameter study and reanalyzed the results.  Because of their higher Reynolds numbers, these new simulations were substantially more computationally expensive than the simulations presented in~\citet{hicks2015} and required a separate computational campaign.  In Section 2, we describe the problem formulation and the setup for the new simulations.   Next, in Section 3, we briefly discuss the turbulent flame regimes and the flame width measurements for the new simulations, which show that these higher $Re$ flames are still thin rather than thick.  Then, in Section 4, we describe the measurements of the flame speed and turbulent velocity for the new simulations and compare these measurements with the predictions of both types of subgrid model.  We show that the flame speed curve on the burning velocity diagram becomes more concave up, so the Turb-SGS model scalings still don't predict the data well.  Then, we show that the RT-SGS model scaling significantly underpredicts the flame speed of the new simulations.  Finally, we discuss some conclusions in Section~5.

\section{Problem Formulation}
\begin{figure*}
\begin{center}
\includegraphics[height=9in,angle=0]{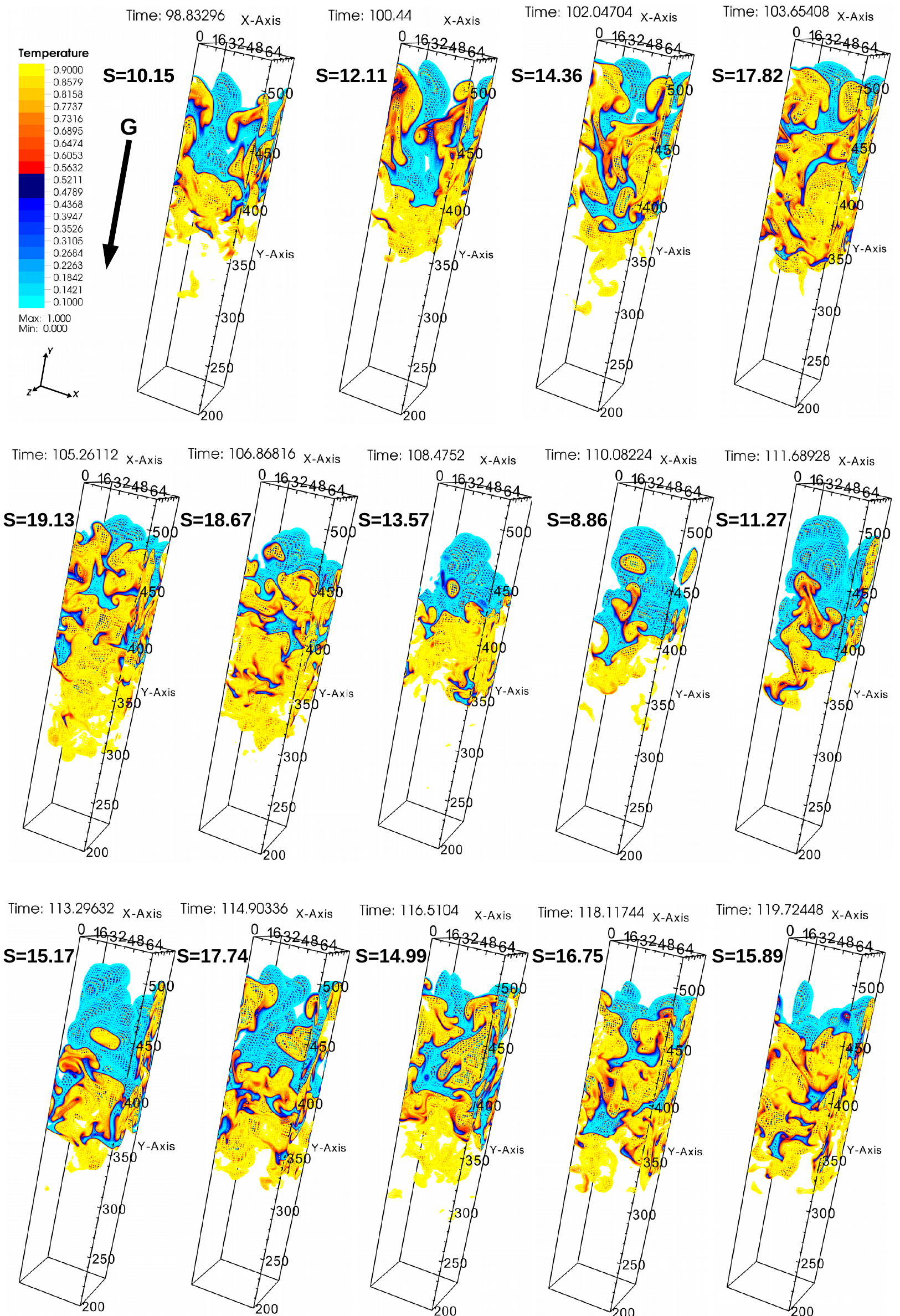}
\end{center}
\caption{Contour Plots of Temperature, Simulation A. Blue colors represent mostly unburnt fuel and red/yellow colors represent mostly burned ashes.  Each flame is propagating in the $y$-direction, against the force of gravity which points in the $-y$ direction.  The instantaneous turbulent flame speed is given to the left of each plot.  See Figure~\ref{fig-4200vs4210} for a plot of the flame speed vs. time.}
\label{fig-flamefront}
\end{figure*}

\begin{table*}
\begin{tabular}{llllllllll}
 \hline
 \hline
     label & G & L & physical size & elements & order & DOF & resolution & time & time step ($10^{-3}$) \\
 \hline
      A & 16   & 64 & 64 x 832 x 64 & 32 x 416 x 32 & 9 &  310,542,336 & 0.222 & 126.96 & 1.728   \\ 
      B & 16   & 64 & 64 x 832 x 64 & 32 x 416 x 32 & 7  & 146,112,512 & 0.286 & 126.96 & 1.728   \\
 \hline
\end{tabular}
\caption{Simulation Parameters.  The columns are: label for simulation, the nondimensional gravity, the
nondimensional domain size, the physical size, the number of elements ($N_x$ x $N_y$ x $N_z$), the
polynomial order ($p_o$), the number of degrees of freedom ($ \sim N_x N_y N_z p_o^3$), the average
resolution (the average spacing between collocation points), the total running time, the
time step.  All quantities are in nondimensional units.}
\label{table-sims}
\end{table*}

In this section, we explain our problem formulation in two phases.  In Section 2.1, we give a straightforward description of our simulation setup which is based on several key simplifications, including the use of the Boussinesq approximation and a model reaction. In Section 2.2, we explore these simplifications in greater detail and provide two Type Ia-relevant interpretations of our nondimensional control parameters.

\subsection{Simulation Setups}

To see whether the conclusions from~\citet{hicks2015} still hold when the turbulence generated by the RT instability is stronger, we added two new simulations to our parameter study.   Both simulations follow the same basic setup as the other simulations in the parameter study.  In this section, we will briefly summarize this setup; for the full details see~\citet{hicks2015}.   

To isolate the effects of the RT instability on the flame, we made two major simplifications to more realistic treatments of nuclear burning.  First, we took advantage of the fact that the density jump across the flame surface is small and simplified the fluid equations using the Boussinesq approximation.  Second, we used a simple model reaction to replace the complex details of nuclear burning.   Specifically, we chose the bistable reaction, $R(T) = 2 \alpha T^2 (1-T)$, with zero ignition temperature and added this reaction term to the advection-diffusion-reaction (ADR) temperature equation.   $T$ is then just a reaction progress variable  that tracks the state of the fluid from unburnt fuel at $T=0$ to burnt ashes at $T=1$.  The bistable reaction type has a simple laminar solution, with laminar flame speed $s_o = \sqrt{\alpha \kappa}$ and laminar flame width $\delta=\sqrt{\dfrac{\kappa}{\alpha}}$, where $\alpha$ is the laminar reaction rate, and $\kappa$ is the thermal diffusivity~\citep{xin2000,vladimirova2003a}.   A better measure of the actual flame width is the thermal flame width, $\delta_T=4 \delta$ (see Section 3).

Non-dimensionalizing the Boussinesq equations with the laminar flame thickness ($\delta$) and the reaction time ($1/\alpha$) gives
\begin{subequations}
\begin{gather}
 \dfrac{D \textbf{u}}{Dt}=-\left(\dfrac{1}{\rho_o}\right)\nabla p + G\,T + Pr
\nabla^{2} \textbf{u} \label{NNS}\\
\nabla \cdot \textbf{u}=0  \label{Nincomp}  \\
\dfrac{D T}{D t} = \nabla^{2} T + 2T^{2}(1-T). \label{Nheat}
 \end{gather}
\end{subequations}
with two control parameters:
\begin{align}
G &=g\left(\frac{\Delta \rho}{\rho_o}\right)\dfrac{\delta}{s_o^{2}} \\
Pr &=\frac{\nu}{\kappa}
\end{align} 
where $G$ is the non-dimensionalized gravity and $Pr$ is the Prandtl number.  The third control parameter is the non-dimensional domain width, $L=\dfrac{\ell}{\delta}$, where $\ell$ is the dimensional length in the $x$ and $z$ directions. We calculate the Reynolds number $Re = u' L$ (when $Pr=1$) from the root-mean-square (rms) velocity measured in the flow (see Section 4.1).  Both new simulations presented in this paper have $L=64$, $G=16$, $Pr=1$.  The parameters for the other simulations are listed in Table 1 of~\citet{hicks2015}. 

All simulations in the parameter study were in 3D with the flame propagating in the $y$-direction against a gravitational force in the $-y$ direction (see Figure \ref{fig-flamefront}).   The boundary conditions were periodic on the sidewalls, inflow into the top of the box, and outflow from the bottom of the box.  The initial flame front was a perturbed plane, with a $tanh$-shaped temperature profile.  The initial velocity was zero throughout the domain.  

The details of the two new simulations with parameters $L=64$, $G=16$ are listed in Table~\ref{table-sims}.   Both simulations were run using \textsc{Nek5000} \citep{nek5000}, a freely-available, open-source, highly-scalable spectral element code currently developed at Argonne National Laboratory (ANL).  Both simulations used 425,984 spectral elements.  Of the two simulations, Simulation A is more highly resolved.   The simulations were run on 32,768 processes on ANL's Mira supercomputer for a total computational cost of several million hours.

Each simulation begins with an initial transient, during which the flame speed grows.  After the transient growth is complete, the flame speed oscillates around a statistically steady average.  Both simulations were run for long enough for the flame speed to undergo several oscillations of its dominant period (see Figure~\ref{fig-4200vs4210}).  Averaged quantities were computed over this oscillating state and ignored the initial transient.  Both new simulations were resolved; the average resolution for both A and B is smaller than the viscous scale calculated from $Re$.   In addition, the resolution is also smaller than the three directional viscous scales.   Finally, the time-averaged flame speeds computed for the two simulations, $s_A=13.86$ and $s_B=13.12$, agree to within six percent.  We consider this adequate, especially given the large oscillations of the flame speed.   Throughout the paper, both simulations are shown in the figures and should be thought of as two different realizations of the flame behavior at $L=64$, $G=16$.

\subsection{Discussion of the Model Assumptions and the Nondimensional Parameters }
As Section 2.1 showed, we chose to make several simplifying assumptions in formulating our simulation setup.  In this subsection, we examine those assumptions in more detail and explain what effects they omit and how they could break down.  In addition, we provide two different dimensional interpretations of the nondimensional control parameters. 

\subsubsection{The Boussinesq Approximation}
One of the goals of this paper is to measure the effect of the Rayleigh-Taylor instability on the flame front.  To isolate the effect of the RT instability, we use the Boussinesq approximation of the full fluid equations.  The Boussinesq approximation is valid for subsonic flows in which density and temperature variations are small~\citep{spiegel1960}.  In this case, the continuity equation is incompressible and density differences in the flow only appear in the gravity-dependent buoyancy forcing term in the Navier-Stokes equation.  Our use of the Boussinesq approximation means that our simulations cannot capture certain effects. First, there is no change in velocity across the flame front due to expansion.  This means that our flames are not susceptible to the Landau-Darrieus (LD) instability.  Second, our simulations cannot include shocks or pressure variations within the flame front.  Third, there is no heating due to the viscous dissipation of energy.  The elimination of these effects is desirable for this study because it simplifies the problem and allows us to study the RT instability in isolation.  

However, the use of any simplifying assumption, like the Boussinesq approximation, generates two questions. First, how good is the assumption? Second, how important are the effects being left out and are they likely to change our results?  Beginning with the first question, the fitness of the Boussinesq approximation depends on the size of the density drop across the flame front. At high fuel densities, this density drop is small; for example, $\Delta\rho / \rho =0.094$ for fuel with composition $X(^{12}C)=0.5$, $X(^{16}O)=0.5$ (ie. a 50/50 CO flame) at $\rho=10^{10} g/cm^3$ ~\citep{timmes1992}.  However, at low densities the density drop is larger, $\Delta\rho / \rho =0.504$ for fuel with the same composition at $\rho=10^{7} g/cm^3$~\citep{timmes1992}, and the Boussinesq approximation is no longer strictly valid.  This leaves open the possibility that, when our simulations are interpreted to represent flames at low density, important effects may be left out.  For example, the velocity induced by the expansion across the flame front might be important, yet ignored.  For the lowest density flames, this velocity will be of order $v\approx s_o$.  Whether this is significant depends on the velocities generated by the Rayleigh-Taylor instability.  We will compare these velocities in Section 3 and show that the RT-generated velocities are much larger and therefore that the expansion-induced velocity is unlikely to affect our conclusions.

Second, we must consider the possibility that the Landau-Darrieus instability could compete with the RT instability to set the flame speed at low densities.  However,~\citet{bell2004a} studied the LD instability for low density CO flames and found a maximum flame speed increase of $s/s_o=1.02$.  \citet{ropke2003} found a flame speed increase of $s/s_o=1.3$.  These flame speed increases are small because LD unstable flames tend to non-linearly stabilize with a low-amplitude cusp shape. Importantly,~\citet{ropke2004a} found ``no convincing indications for active turbulent combustion'', that is, no sign that the LD instability can lead to drastically higher flame speeds, at fuel densities down to $\rho=10^{7} g/cm^3$. ~\citet{ropke2004b} reached a similar conclusion for LD unstable flames propagating into a vortical fuel.  The flame speed increase for the RT unstable simulations in this paper is much larger, $s_A=13.86$ and $s_B=13.12$, suggesting that the RT instability overwhelms the LD instability.  

Finally, our simulations cannot capture pressure variations within the flame front, shocks or detonations.  In particular, they are not able to probe the kind of flame self-acceleration DDT mechanism described by~\citet{poludnenko2011,poludnenko2015,poludnenko2017_icders} and recently extended to turbulent $^{12}C$ flames~\citep{poludnenko2019_aas}.  Whether a similar mechanism could lead to the detonation of RT unstable flames propagating into a quiescent fuel is an interesting question which is mostly beyond the scope of this paper.

\subsubsection{Flame Reaction Model}

\begin{figure}
\begin{center}
\includegraphics[height=3.5in,angle=270]{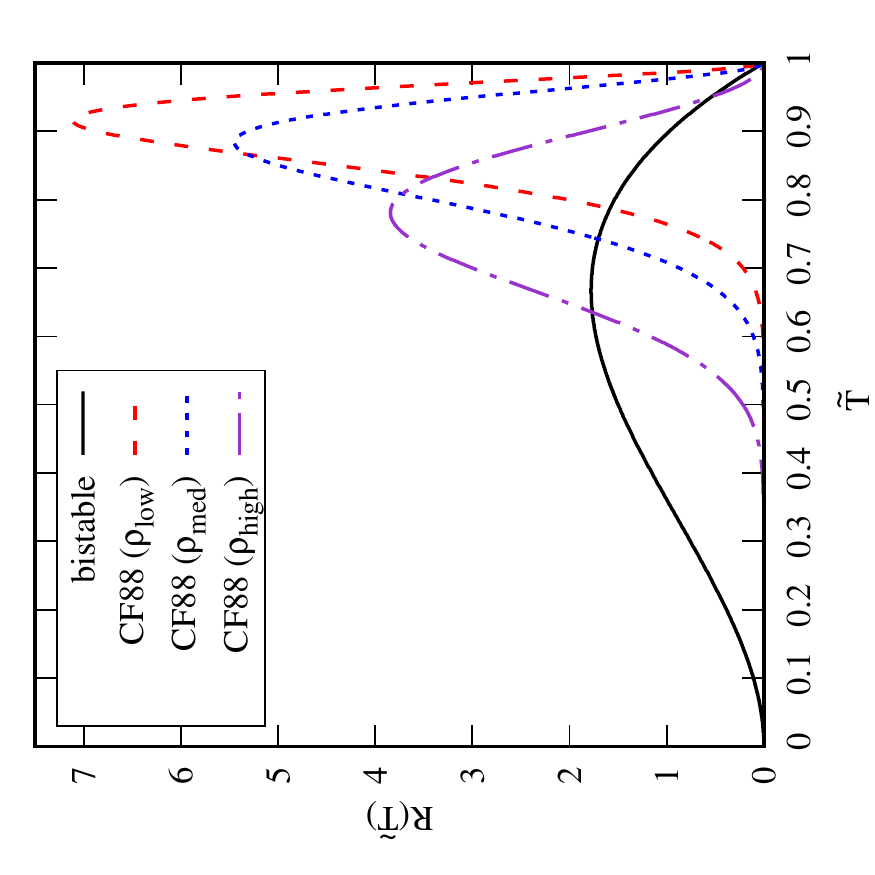}
\end{center}
\caption{Reaction Models Comparison.  This figure shows a comparison between the bistable reaction (solid black curve) and the CF88 reaction at three different densities, $\rho_{\rm low}=6.67\times10^6 g\,cm^{-3}$ (long-dashed red curve), $\rho_{\rm med}=2\times10^8 g\,cm^{-3}$ (short-dashed blue curve), and $\rho_{\rm high}=10^{10} g\,cm^{-3}$  (dot-dashed purple curve).  $R(\widetilde{T}$) is the nondimensionalized reaction rate, normalized so that the area under each curve is $1$.  $\widetilde{T}$ is a progress variable that goes from $0$ (unburned) to $1$ (burned).  The CF88 reaction rate is more peaked (ie. has a lower FWHM) than the bistable reaction, especially at low density.  Data for this figure are given in Table~\ref{tab-cf88}.}
\label{fig-cf88}
\end{figure}

\begin{table*}
\begin{tabular}{lllllll}
 \hline
 \hline
    reaction & $\rho_{9}$ & $\Delta \rho / \rho_o$  &  $T_{b,9}$ & r & $\widetilde{T}$ at max & FWHM \\
 \hline
     CF88 ($\rho_{\rm low}$) & 0.0067& 0.529 & 2.4 & 1 & 0.916 & 0.133\\
     CF88 ($\rho_{\rm med}$) & 0.2 & 0.2248 & 5.16 & 1 & 0.884 & 0.175\\
     CF88 ($\rho_{\rm high}$) & 10 & 0.094 & 11.12 & 1 & 0.781 & 0.252\\
    bistable & -- & --  & --  & --  & 0.666 & 0.577\\
 \hline
    CF88  & 0.0067& 0.529 & 2.4 & 0.01 &  0.916 & 0.132\\
    CF88  & 0.0067& 0.529 & 2.4 &  0.1&  0.915 & 0.133 \\
    CF88  & 0.0067& 0.529 & 2.4 &  10&  0.933& 0.116\\
    CF88  & 0.0067& 0.529 & 2.4 &  100&  0.974& 0.0767\\
    CF88  & 0.0067& 0.529 & 2.4 &  1000&  0.995& 0.0577\\
 \hline
\end{tabular}
\caption{Flame Reaction Rate Comparison Data for Figure~\ref{fig-cf88}. The columns are the reaction type, the fuel density divided by $10^9 \rm g\,cm^{-3}$, the density drop across the flame, ash temperature divided by $10^9 \rm K$, the scaling exponent for the density model, the measured value of the progress variable at which the peak reaction rate occurs, and the measured FWHM of the reaction rate.  The top half of the table gives the data for the reactions shown in Figure~\ref{fig-cf88}.  The bottom half of the table shows the dependence of $\widetilde{T}$ at the peak and the FWHM on $r$ for the lowest density reaction.  The flame data are from \citet{bell2004b} (Table 1, and pg 902) for $\rho_9=0.0067$ and from \citet{dursi2003} (Table 3) for $\rho_9=0.2$.  The density drop data for $\rho_9=10$ is from~\citet{timmes1992} (Table 1).  We calculated the ash temperature for $\rho_9=10$ using an isochoric self-heating calculation for a 50/50 CO mixture as described by~\citet{calder2007} using the Timmes EOS~\citep{timmes1999,timmes2000a} downloaded from  \url{http://cococubed.asu.edu/code_pages/eos.shtml}.  All data are for 50/50 CO flames.}
\label{tab-cf88}
\end{table*}

The second major simplification in our investigation is the use of a simple flame model.  In reality, nucleosynthesis in Type Ia supernovae is a complex process involving hundreds of nuclear species~\citep{seitenzahl2017}.  This burning process can be divided into three basic stages: carbon burning to magnesium, burning of oxygen, neon and magnesium to intermediate mass elements like silicon, and, finally, burning to heavy nuclei like iron~\citep{khokhlov2000,calder2007, townsley2007}. These stages have well-defined timescales, with carbon burning happening the fastest, especially in low density fuel.  Several studies of flame dynamics in Type Ia supernovae have taken advantage of this time scale separation to justify considering only carbon burning in their simulations~\citep{bell2004a,bell2004b,bell2004c,zingale2005a}.  They model this single reaction using the reaction rate from~\citet{caughlan1988}, which is often referred to as the CF88 reaction. The goal of this subsection is to compare the simple model reaction that we use (the bistable reaction) to the CF88 reaction.

The Boussinesq equations and the fully compressible fluid equations are different, so comparing the reaction rate models used in their ``reaction terms'' requires some thought.  First, it must be clear what we mean by ``reaction rate''.   Typically, the term ``reaction rate'' refers to the temperature-dependent part of the reaction, but in some papers (for example,~\citet{dursi2003}) it refers to the net rate of change of the abundance of a species due to nuclear reaction.  In this paper, we use the second meaning so that the CF88 reaction rate is given by
\begin{equation}
\frac{dX_C}{dt} = -\frac{1}{12}X_C^2 \, \rho \, r(T),
\end{equation}  
with a temperature dependence of 
\begin{equation}
\begin{aligned}
r(T) =  & (4.27 \times 10^{26}) \frac{T_{9,a}^{5/6}}{T_9^{3/2}} \\
           & \times exp \left( \frac{-84.165}{T_{9,a}^{1/3}} - (2.12\times 10^{-3}) T_9^3  \right),
\end{aligned}
\end{equation}  
where $T_9=T/10^9$ K, $T_{9,a} = T_9 / (1+0.0396 T_9)$ and $X_C$ is the mass fraction of $^{12}C$~\citep{caughlan1988}.  For this subsection only, $T$ refers to the actual dimensional temperature in K.  Elsewhere in the paper, $T$ is the non-dimensional reaction progress variable which goes from $0$ to $1$ during the burning process.  In this subsection, this progress variable is instead represented by $\widetilde{T}$.

In the compressible case, the reaction rate for each species appears in the energy equation and in the advection equation for that species.  Applying the Boussinesq approximation, considering the case when thermal and species diffusion are equal ($Le=1$), nondimensionalizing, and rescaling so that $\widetilde{T}$ is a progress variable that goes from 0 to 1, collapses these equations into the single temperature evolution equation that we use here, Eqn.~\ref{Nheat}.  In this equation, the reaction rate is given by the term $R(\widetilde{T})= 2 \widetilde{T}^2 (1-\widetilde{T})$.  Comparing the Boussinesq and compressible forms of the fluid equations suggests that the best ``reaction rate'' comparison is between $R(\widetilde{T})$ and $|dX_C / dt|$ because they play analogous roles in their respective equations and because they are both zero (or nearly zero) in pure fuel and ashes and have a peak reaction rate at some mixture of fuel and ashes.  In contrast, $r(T)$ alone has a peak value at a temperature much higher than the temperature of the ashes, so it cannot be consistently compared with $R(\widetilde{T})$.  

In order to compare the CF88 reaction with the bistable reaction, it must be transformed to depend only on the progress variable $\widetilde{T}$.  First, $\rho$ and $X_C$ must be expressed in terms of $\widetilde{T}$.   The relationship between $\rho$ and $T$ is mediated by the fluid equations and the equation of state (EOS), so it is not possible to derive a simple expression for $\rho(\widetilde{T})$ directly from first principles.  Therefore, we use a model expression $\rho=\rho_o[1-(\Delta \rho / \rho_o) \widetilde{T}^r]$, where $\rho_o$ is the density of the fuel and $\Delta\rho=\rho_o-\rho_{ash}$, and then test the sensitivity of the result to the power law parameter $r$.  On the other hand, $X_C$ can be directly expressed in terms of $\widetilde{T}$.  First note that $X_{C,mix} = (1-f)  X_{C,fuel}$ for a mixture of fuel and ash, where $f$ is the mass fraction of ash.  Since $T_{fuel} << T_{ash}$, $f=(C_P(T_{mix}) T_{mix})/(C_P(T_{ash}) T_{ash})$ so determining the $T$ dependence of $X_C$ reduces to determining the $T$ dependence of $C_P$, the specific heat at constant pressure~\citep{bell2004b}.  The specific heat includes contributions from ions, electrons, and radiation and it depends on both $T$ and $\rho$.  Rewriting the expression for $C_P$ from~\citet{woosley2004} (their Eqn. 6) in terms of $\widetilde{T} = T/T_{b,9}$ (where $T_{b,9} = T_{ash}/(10^9 K)$ is the burned temperature) gives  
\begin{equation}
\begin{aligned}
C_P=\, 9.1\times10^{14} + & \frac{(8.6\times10^{14})\,\widetilde{T}\,T_{b,9}}{\rho_9^{1/3}} \\
                                         + & \frac{(3.0\times10^{12})(\widetilde{T}\, T_{b,9})^3}{\rho_9}
\end{aligned}
\end{equation}  
with units of ergs $\rm g^{-1}$ $(10^8 \rm K)^{-1}$.  Combining the density model and this expression for $C_P$ yields an expression for the mass fraction of ash, $f=f(\widetilde{T}, T_{b,9}, \rho_{o}, \Delta \rho / \rho_o,  r)$.  Using all of these elements, we can then construct an equation for the CF88 reaction rate that depends only on the progress variable $\widetilde{T}$, but takes the ash temperature ($T_{b,9}$), the fuel density ($\rho_{o}$), the density jump across the flame ($\Delta \rho / \rho_o$), and the scaling for the density model ($r$) as input parameters.   These input parameters are not independent; a given initial density and temperature will result in a predictable density jump and final ash temperature.  However, we use tabulations of these values from other sources (or calculate them) and treat them as independent inputs into the CF88 renormalized expression.

The resulting comparison between the bistable reaction and CF88 reaction is shown in Figure~\ref{fig-cf88}. The CF88 reaction is shown for low, medium, and high density fuels.  Each density has a different sets of inputs, given in Table~\ref{tab-cf88}.  The reaction curves in the figure are normalized so that the area under each curve is 1.  To compare the reactions, we will look at two quantities that describe their shapes: the value of $\widetilde{T}$ at which the peak reaction rate occurs and the peakedness of the reaction as described by the full width at half max (FWHM) of the curve.  Looking at the figure, it is clear that the CF88 reaction is always more highly peaked than the bistable reaction and that this peak occurs at higher $\widetilde{T}$ (for measurements of the FWHM and $\widetilde{T}$ at the peak see Table~\ref{tab-cf88}).  The CF88 reaction is most highly peaked for fuel at low density, and least highly peaked for fuel at high density.  The $\widetilde{T}$ at which the peak reaction occurs is highest for the fuel at low density. The parameter $r$ does not have much effect on the CF88 reaction shape for the highest density fuel for values of $r$ from 0.01 to 1000.  For medium density fuel, the FWHM of the reaction curve stays roughly the same while the $\widetilde{T}$ at the peak of the curve goes from $0.884$ at $r=0.01$ to $0.925$ at $r=1000$.  For the lowest density fuel, increasing $r$ decreases the FWHM and increases the $\widetilde{T}$ at the peak of the curve (see Table~\ref{tab-cf88}).  Overall, the bistable reaction is a better approximation for fuel at high density than fuel at low density, but it is always substantially wider than the CF88 reaction.  Potential effects of the reaction type on our results will be discussed later in Sections 3, 4 and 5.

\subsubsection{Interpreting the Non-Dimensional Parameters}
In Section 2.1, we formulated the problem in terms of the dimensionless variables $G$, $L$, and $Pr$.  Working with non-dimensional variables is a powerful approach because it allows for generalization.  However, a lack of connection with the dimensional quantities can lead to a sense of unreality.  In addition, it may be less obvious when the approximations that make the nondimensional approach possible have broken down.  Here, we provide two dimensional interpretations of our dimensionless variables, though many other interpretations are possible.

The first interpretation is the most straightforward: our goal is to match the dimensional flame properties to the nondimensional variables as closely as possible.  In this ``matched flame" interpretation, the simulation domain represents a tiny physical box with a realistically sized flame inside.  The one choice that must be made is how to match the measured flame width of a real flame with a dimensional version of our laminar flame width, $\delta$.  If the simulated bistable  flame represents the reaction zone of the highly peaked real flame, then it would be reasonable to set $\delta = l_F / \delta_T$, where $l_F\equiv(T_{ash}-T_{fuel})/max( \nabla T)$ is the dimensional thermal width of the real flame and $\delta_T=4$ is the dimensionless thermal width of the bistable flame.  Alternatively, if the simulated flame represents the entire flame then it would be reasonable to set $\delta = l^{0.9}_{0.1} / \delta^{0.9}_{0.1}$, where $ l^{0.9}_{0.1}$ is the dimensional width of the real flame measured from $10\%$ above the fuel temperature to $90\%$ of the ash temperature and $\delta^{0.9}_{0.1}=4.394$ is the dimensionless width of the bistable flame measured between progress variable values $T=0.1$ and $0.9$.  Then, the matched $\delta$, $g$, and the flame properties, $\Delta\rho/\rho$ and $s_o$, can be combined to find $G$ for a flame at any given density.  For example, using $g=10^9 \rm cm\,s^{-2}$, $\Delta\rho/\rho=0.504$, $l^{0.9}_{0.1}=4.22\, \rm cm$, and $s_o=4.73\times10^3 \rm cm/s$  (\citet{timmes1992}, Table 3 properties for a 50/50 CO flame at $\rho=10^7 \rm g/cm^3$, calculated using a 130 isotope network) and the choice that the bistable flame represents the entire flame, we find $G=21.6$.  Or, using $g=10^9 \rm cm\,s^{-2}$, $\Delta\rho/\rho=0.482$, $l_T=1.9\, \rm cm$, and $s_o=2.97\times10^3 \rm cm/s$  (\citet{bell2004b}, Table 1 properties for a 50/50 CO CF88 flame at $\rho=10^7 \rm g/cm^3$)  and the choice that the bistable flame represents the reaction zone, we find $G=26.0$.  Both of these values are relatively close to our simulated value, $G=16$, so according to the matched flame interpretation, the simulated flame represents a flame at a density slightly higher than $\rho=10^7 \rm g/cm^3$, in a physical domain tens of cm in width and several meters in height.  In this interpretation, $G$, $L$ and $GL$ match the real flame and there is a straightforward translation of length and time scales using $\delta$ as the length scale and $\delta / s_o$ as the time scale.  For example, the dimensional value of the rms velocity will be $u' s_o$.  However, the $Re$ measured from these $Pr=1$ simulations will be smaller than the actual $Re$ by a factor of $1/Pr\approx 10^5$.  Overall, the matched flame interpretation is self-consistent and straightforward, but it ignores all scales larger than the box size.

The second interpretation is that the simulated flame represents a thickened flame in a subgrid scale (SGS) sized box.  In this ``thickened flame" interpretation, our goal is to match $GL$, which measures the importance of the RT instability relative to laminar burning.  (Note that the densimetric Froude number is $Fr_d=1/\sqrt{GL}$.)  Choosing a value for $GL$, $g$, and the flame parameters, $\Delta\rho/\rho$ and $s_o$, we can calculate the dimensional subgrid scale that simulation domain represents using $l_{SGS} = (G L \, s_o^2) / (g \, \Delta\rho/\rho)$.  Using the tabulated data in \citet{timmes1992} for 50/50 CO flames and $g=10^9 \rm cm\,s^{-2}$, we find that our $GL=1024$ simulations could represent flames with, for example, $(\rho_9,l_{SGS}) = \rm (0.05, 76\,m), (0.1,1.3\,km), (0.2, 9.3\,km), (0.5, 139\,km)$.  In this interpretation, although the product $GL$ is matched, $G$ is too large and $L$ is too small because $\delta$ is the laminar flame width of the thickened flame, not the real flame.  However, velocities scale correctly because the thickened flame travels at the correct laminar flame speed, so $u' s_o$ is the dimensional rms velocity.  Once again, the $Re$ is too small.  Overall, the thickened flame interpretation is less straightforward, but it provides the necessary bridge between the matched flame interpretation and the development of subgrid scale flame speed models.   

The final nondimensional parameter in our simulations is the Prandtl number, the ratio of the diffusion of momentum to the diffusion of heat.  The $Pr$ in the star is very small, $\sim10^{-5}$~\citep{timmes1992}, because the thermal diffusivity is much larger than the viscosity. Scalings for these quantities are given by~\citet{nandkumar1984}.  Both viscosity ($\nu$) and thermal diffusivity ($\kappa$) vary with density and $\kappa$ is also dependent on temperature.  This means that $Pr$ varies across the flame front.  In addition, the temperature dependence of $\kappa$ influences the temperature structure of the laminar flame front.  This is another factor that makes the structure of CF88 flames different from the structure of bistable flames.  

Small $Pr$ fluids are hard to simulate because they have a wide separation between the viscous and thermal scales so a higher resolution is required.  Research on Type Ia flames has dealt with this problem in two ways.  First, because the viscosity is so small it is reasonable to ignore it entirely and simulate the inviscid Navier Stokes equations or inviscid low Mach number equations~\citep{bell2004a,bell2004b,bell2004c,zingale2005a}.  In this case, turbulent eddies are dissipated at small scales by the intrinsic numerical viscosity in the simulation.  An effective $Re$ can be estimated from these types of simulations, but there are some uncertainties involved~\citep{aspden2008a}.   In this paper, we use a second approach which is to include viscosity and set $Pr=1$.  We did this because we are specifically interested in the interaction between the turbulence generated by the RT instability and the flame front and so we want to resolve the entire turbulent cascade, down to the Kolmogorov scale.  This eliminates any uncertainty about whether small eddies that could affect the flame are being captured.  We will explore the effect of the $Pr$ in future work.  

Given the interpretational complexities introduced by our ``simplifying'' assumptions and our use of nondimensional control parameters, it is fair to ask whether this approach is worth the trouble. Why not simply simulate realistic CF88 flames using a low Mach or even a compressible code as others have done?  After all, there are clear advantages to realism.  However, there are also some benefits to our complementary approach.  First, by using the Boussinesq approximation and the relatively thick, more easily resolved bistable flame we reduced our computational costs and could carry out a large parameter study (~\citet{hicks2015} and the two simulations in this paper) and run those simulations for long enough to get the robust time averages needed to test the flame speed model predictions.  Second, we could strip out all effects except for the ones we wanted to study: the RT instability, turbulence generated by the RT instability, and burning.  This is useful because the ideas and models that we wanted to test only depend on those three effects.  For example, the hypothesis that RT unstable flames should be thickened by their self-generated turbulence and thereby transition from flamelets to thin reaction zones only depends on the presence of the RT unstable flame and the turbulence it generates. Likewise, the hypothesis that the flame speed should be set by the turbulent velocity again depends only on the presence of the RT unstable flame and turbulence.  Other effects are extraneous in the sense that they are not invoked by these particular hypotheses.  Of course, that is not to say that omitted effects, like compressibility, are necessarily unimportant.  In fact, if there is a flame-driven DDT, compressibility effects must be critically important!  But we are testing a different hypothesis here.  Finally, we believe that there is a benefit to starting with the simplest version of a complex problem, understanding that, and then adding complicating effects to see how they change the picture.       

\section{Turbulent Flame Regimes and the Flame Width}

\begin{figure}
\begin{center}
\includegraphics[height=3.5in,angle=270]{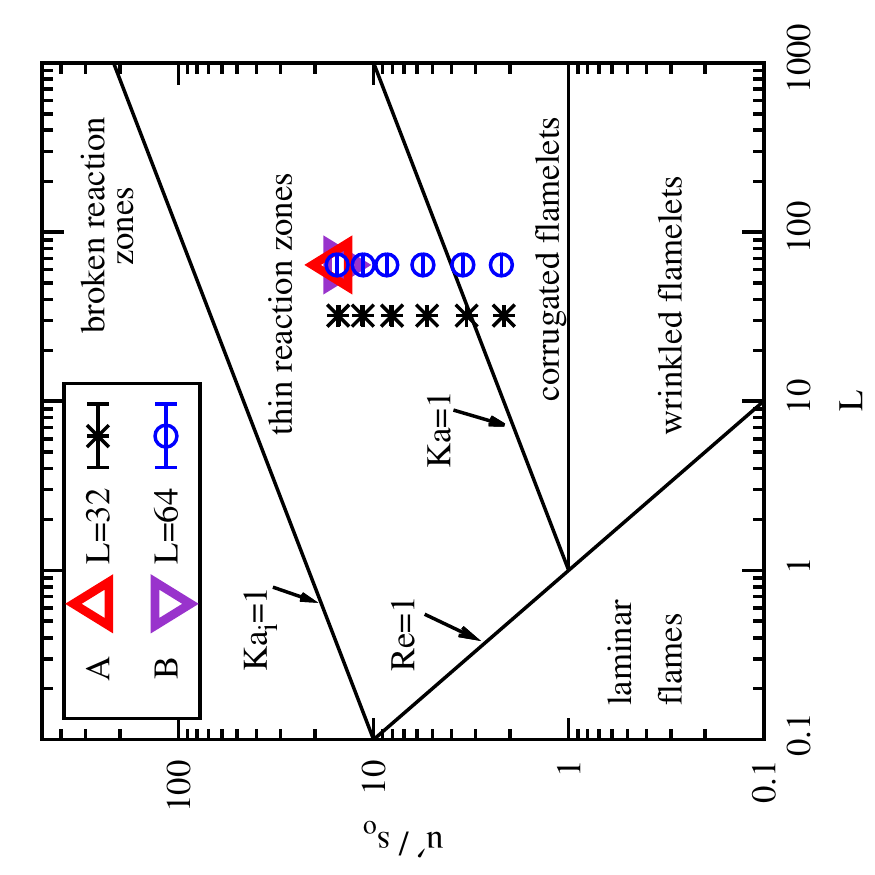}
\end{center}
\caption{Combustion Regimes Diagram.  Predicted divisions between different types of flame behavior are shown with solid lines (adapted from \citet{peters2000}).   Regimes are based on comparisons between the time, velocity, and length scales of turbulence and a laminar flame.  Here, $Re=u' L$, $Ka = (\delta / \eta)^2$, and $Ka_i =(\delta_i / \eta) ^2$, where $\delta_i$ is the width of the innermost reaction zone; we assume $\delta = 10 \delta_i$.   Simulations are shown by black asterisks ($L=32$) and blue circles ($L=64$).  The two new simulations at $L=64, G=16$ are shown by a red triangle (Simulation A) and an inverted purple triangle (Simulation B).  Most of the simulations, including A and B, are predicted to be in the thin reaction zones regime and should have thickened flames.}
\label{fig-regimes}
\end{figure}

\begin{figure*}
\begin{center}
\includegraphics[height=6in,angle=270]{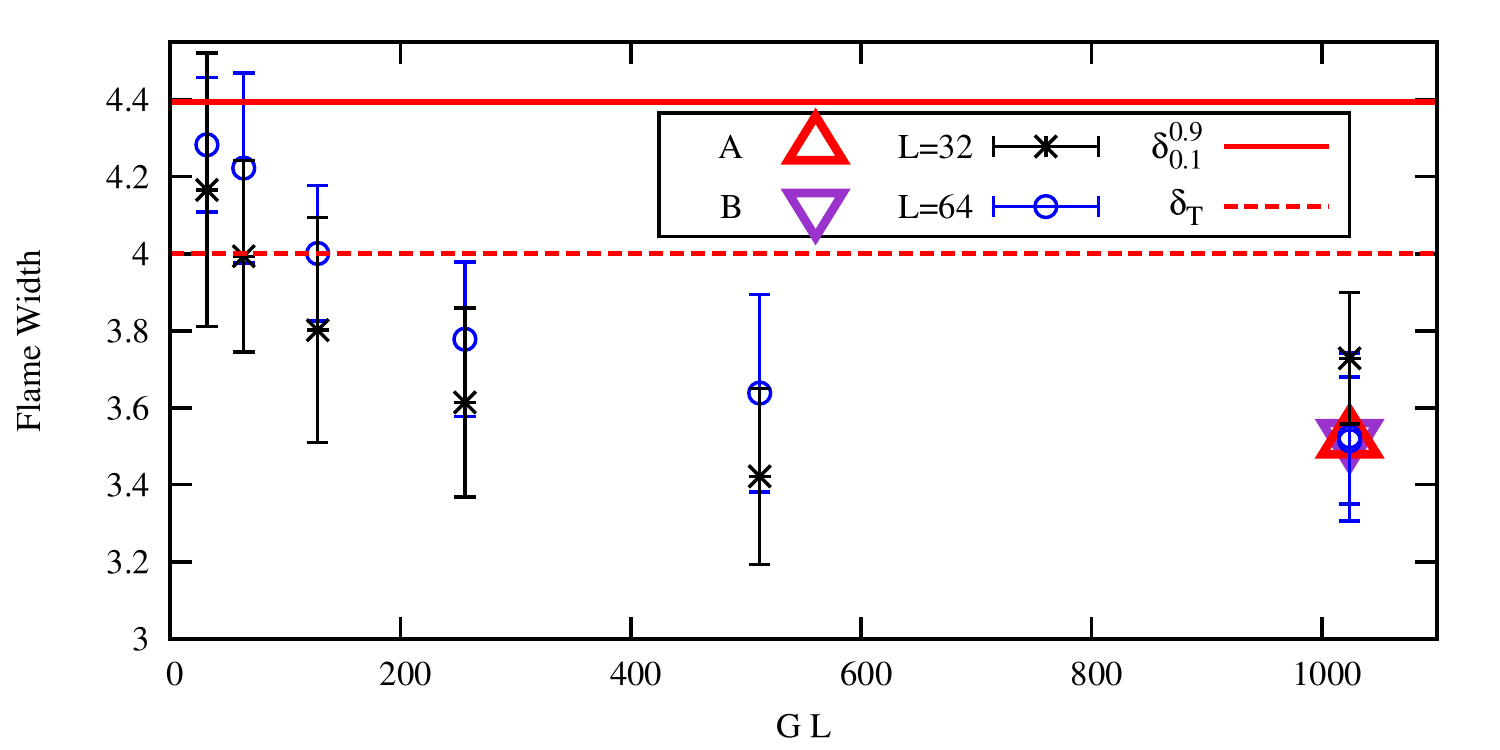}
\end{center}
\caption{Turbulent Flame Width vs. GL.  For each simulation, the time-averaged flame width between the contours $T=0.1$ and $0.9$ was calculated in post-processing using the iterative isosurface-based method  (see~\citet{hicks2015}, Section 3) .   Error bars show the minimum and maximum flame widths calculated using this method.  The laminar flame width between $T=0.1$ and $0.9$, $\delta^{0.9}_{0.1} = 4.394$ , is the solid red line.  The laminar thermal flame width, $\delta_T =4$, is the dashed red line.  Surprisingly, most of the simulations, including A and B, have a flame width smaller than $\delta^{0.9}_{0.1}$, which implies that the flames are stretched flamelets instead of thin reaction zones although $Ka>1$ and $Ka_T >> 1$.}
\label{fig-widths}
\end{figure*} 

In~\citet{hicks2015}, we showed that RT unstable flames don't follow the combustion regime predictions of traditional turbulent combustion theory, which deals with flames burning through upstream turbulence.   Specifically, we found that the flame is thinner than the thermal laminar flame width ($\delta_T = 4$ for a bistable flame), rather than thicker, when turbulence is strong.   The physical mechanisms behind the turbulent combustion regime predictions also underlie the turbulent flame speed models (Turb-SGS), so this result casts doubt on the physical validity of the Turb-SGS models.   Our results also posed a problem for DDT mechanisms that rely on flame thickening to lead to the DDT.   In this section, we show that the two new simulations strengthen these findings.

Turbulent combustion theory divides flame behavior into various regimes based on velocity and length scale ratios of the turbulence and the flame (see Figure~\ref{fig-regimes}).  Two regimes are important here: corrugated flamelets and thin reaction zones. (For a discussion of the other regimes see~\citet{peters2000,hicks2015}.)  Whether a flame is a corrugated flamelet or a thin reaction zone depends on the ratio between the flame reaction time and the eddy turnover time of the viscous scale eddies: $Ka = t_F / t_{\eta}$ (the Karlovitz number), which is equal to the squared ratio of the laminar flame width ($\delta$) and the viscous scale ($\eta$) or the squared ratio of the velocity at the viscous scale ($v_\eta$) and the laminar flame speed ($s_o$), $Ka = (\delta / \eta)^2 = (v_\eta / s_o)^2$, when $Sc = Pr = 1$.   In the corrugated flamelets regime ($Ka < 1$), turbulent eddies wrinkle the flame front, but do not change its internal structure. So, the flame width should be the same as the laminar flame width.  In the thin reaction zones regime ($Ka > 1$), turbulent eddies interact with the internal structure of the flame, increasing the flame width. Specifically, small eddies that ``turn over'' faster than the laminar flame reaction time ($t_F=1/\alpha$) increase the effective thermal diffusivity of the flame, thickening it.

In~\citet{hicks2015} we compared the flame width behavior predicted by the turbulent combustion regime theory to measurements of the flame widths of RT unstable flames.   In order to do this, we had to define Karlovitz numbers for our bistable model flames, which are thicker than the more realistic CF88 model.  We used two definitions:  the standard definition based on the laminar flame width $Ka = (\delta / \eta)^2$, and a thermal Karlovitz number based on the full thermal flame width $\delta_T = 4 \delta$ giving $Ka_T = (4 \delta / \eta)^2$.   $Ka$ is useful for comparing our data with experiments and other simulations, but $Ka_T$ is a better measure of how well eddies penetrate the actual physical flame width.   The predicted flame regimes, based on $Ka$ are shown in  Figure~\ref{fig-regimes}.  Most of the simulations, including A and B, are predicted to be in the thin reaction zones regime.   Regimes based on $Ka_T$ instead would predict these simulations to be even further into the thin reaction zones regime.

Since most of the simulations are predicted to be in the thin reaction zones regime, their flames should be thickened by small eddies.  To check this prediction,~\citet{hicks2015} measured the flame width between the $T=0.1$ and $T=0.9$ contours using the iterative isosurface-based method~\citep{poludnenko2010} and compared the measured widths to the thermal flame width $\delta_T=4$.   We incorrectly stated in the Section 2 of~\citet{hicks2015} that $\delta_T$ is the width between the $T=0.1$ and $T=0.9$ contours.  In fact, the thermal flame width is $\delta_T = (T_{max} - T_{min}) / max(\nabla T) = 1/max(\nabla T)$ for our bistable flame, and the width of the laminar flame measured between the $T=0.1$ and $T=0.9$ contours is $\delta^{0.9}_{0.1} = 4.394$.   So the correct comparison should be between the measured flame widths and $\delta^{0.9}_{0.1}$, which is what we use in this paper.  The change to this comparison does not alter the qualitative conclusions of~\citet{hicks2015}.    

Although most of the simulations in the~\citet{hicks2015} parameter study have $Ka > 1$ (which should place them in the thin reaction zones regime), we found that their flame widths are actually thinner than the thermal laminar flame width (and therefore also thinner than $\delta^{0.9}_{0.1}$).  This is very strange behavior: no turbulent combustion regime predicts a flame width less than $\delta_T$.  The addition of Simulations A and B to the parameter study allowed us to see whether this strange behavior continued at $L=64, G=16$.   It did, as shown in Figure~\ref{fig-widths}.  We found that the flame width for both simulations was about $3.5$, although $Ka=7.5$ and $Ka_T=120$ for A and $Ka=7.3$ and $Ka_T=117$ for B.  Stretching by the RT instability seems to overwhelm any thickening by turbulence, and we still observe no transition to the thin reaction zones regime.

In some ways, it is not surprising that the behavior of A and B is consistent with our previous results.   Although A and B do have the highest $Re$ of any of our simulations so far, they have only the second highest $Ka_T$; the $L=32, G=32$ (L32G32) simulation has $Ka_T = 166$.  If an eventual transition to thin reaction zones is controlled by $Ka$ (as predicted by the turbulent combustion regimes), we would expect to find it at some $Ka_T > 166$, not at $Ka_T=120$.   On the other hand, it is far from clear that $Ka$ is the main parameter controlling the flame width.

The new flame width data from Simulations A and B are consistent with the physical picture of RT unstable flames that we developed in~\citet{hicks2015}.   Vorticity is created baroclinically by horizontal temperature gradients across the flame front of an RT unstable flame, but it is apparently washed downstream fast enough that it doesn't have a significant effect on the flame width.   Unlike turbulent flames, RT unstable flames aren't forced to interact with every turbulent eddy as they propagate. Instead of being thickened by turbulence, the flame is thinned by the stretching action of the RT instability.   RT unstable flames don't fit the physical picture of traditional turbulent flames and don't transition to thin reaction zones, at least for the parameter values that we've studied.  This suggests that Turb-SGS models, which are based on the assumption that RT unstable flames behave like turbulent flames, may not be a physically sensible choice.   These findings are also a problem for DDT models, like the Zel'dovich gradient model, which require a transition to thin reaction zones for a detonation to occur.   The question for future work is whether such a transition will occur at higher values of $Ka$ or $G$ than we've studied so far. 

How might these results be affected by compressibility and reaction type? For flames in low-density fuel ($\rho\sim10^7 \rm g\,cm^{-3}$), gas expansion induces a velocity of $\gamma \, s_o/(1-\gamma) \approx s_o$ when $\gamma=\Delta\rho/\rho \approx 0.5$.  \citet{peters2000} explores the effect that gas expansion has on the turbulent velocity and concludes that it is only important for turbulent flames when the expansion induced velocity is larger than the turbulent fluctuations.  For the simulations in this paper, $Ka>1$ so the turbulent velocity is larger than the induced velocity all the way down to the viscous scale.  This implies that adding compressibility should not affect our results. On the other hand, the response of the flame to RT stretching is likely to depend on the reaction type.  Flames with more peaked reactions (like the CF88 flame) are less susceptible to flame thinning by stretching, but they are also less susceptible to turbulent thickening.  An important question for future work is how these competing effects balance for flames with highly peaked reaction rates.

\section{The Flame Speed and Comparison with Flame Speed Models}

In this section, we will test whether adding the data from the two new simulations changes our main conclusions from~\citet{hicks2015} about the ability of turbulence-based and RT-based flame speed models to predict the flame speed.  In Section 4.1, we begin by presenting the measurements of the flame speed and the rms velocity.   Next, in Section 4.2, we consider turbulence-based models and check our unusual finding that the flame speed data are concave up on the burning velocity diagram (the plot of $s$ vs. $u'$), which suggests that RT unstable flames are fundamentally different from turbulent flames.  We also re-evaluate how well linear, scale-invariant and bending models predict the flame speed.  Finally, in Section 4.3 we will see whether the RT-based model still underpredicts the flame speed at high $GL$. 

\subsection{Measurements of the Flame Speed and Turbulent Velocity}

\begin{figure*}
\begin{center}
\includegraphics[height=6in,angle=270]{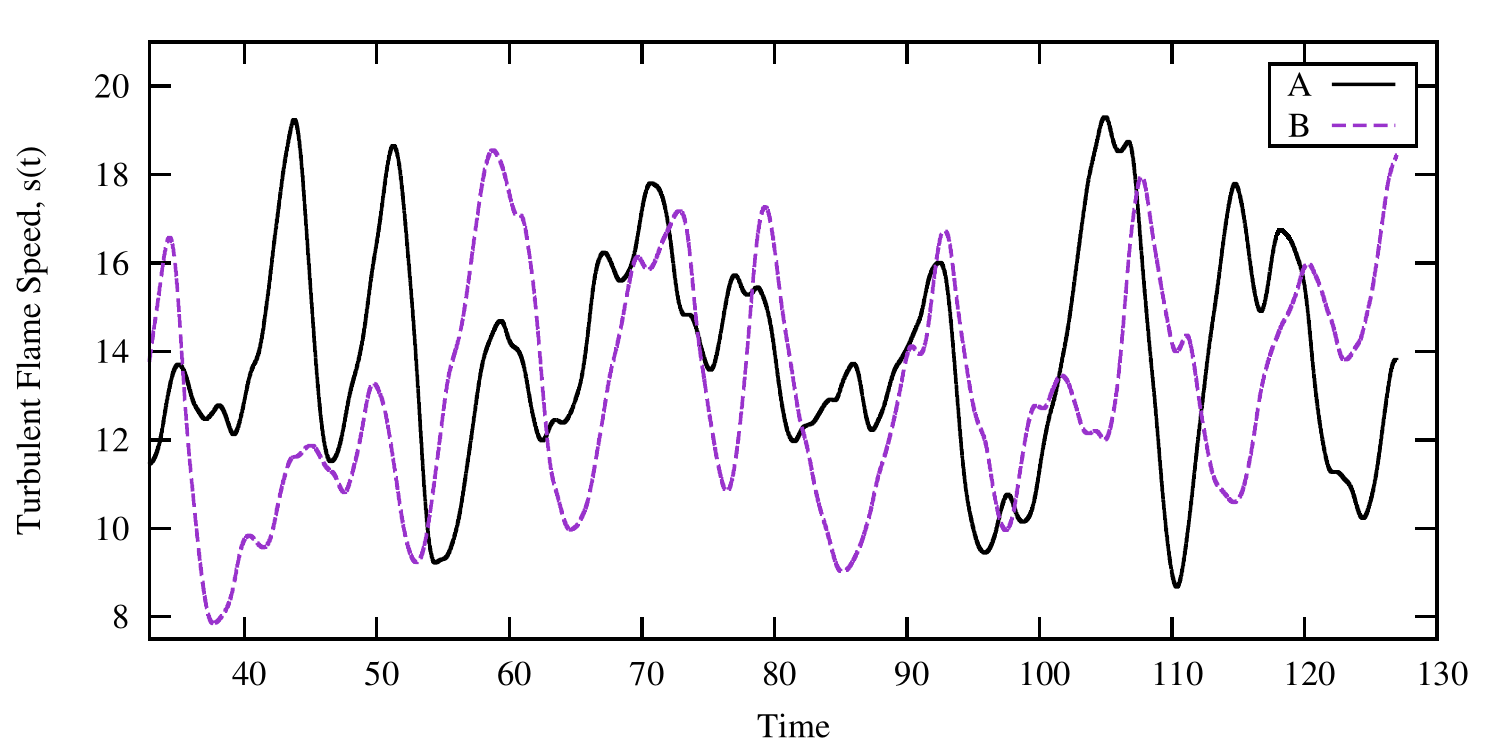}
\end{center}
\caption{Turbulent Flame Speeds vs. Time.  The turbulent flame speed, $s(t)$, for Simulations A (solid black line) and B (dashed purple line) is shown after it has reached saturation; it is during this period of time that the average flame speed is calculated.  Note the large amplitude of the flame speed oscillations.}
\label{fig-4200vs4210}
\end{figure*}

\begin{figure*}
\begin{center}
\includegraphics[height=6in,angle=270]{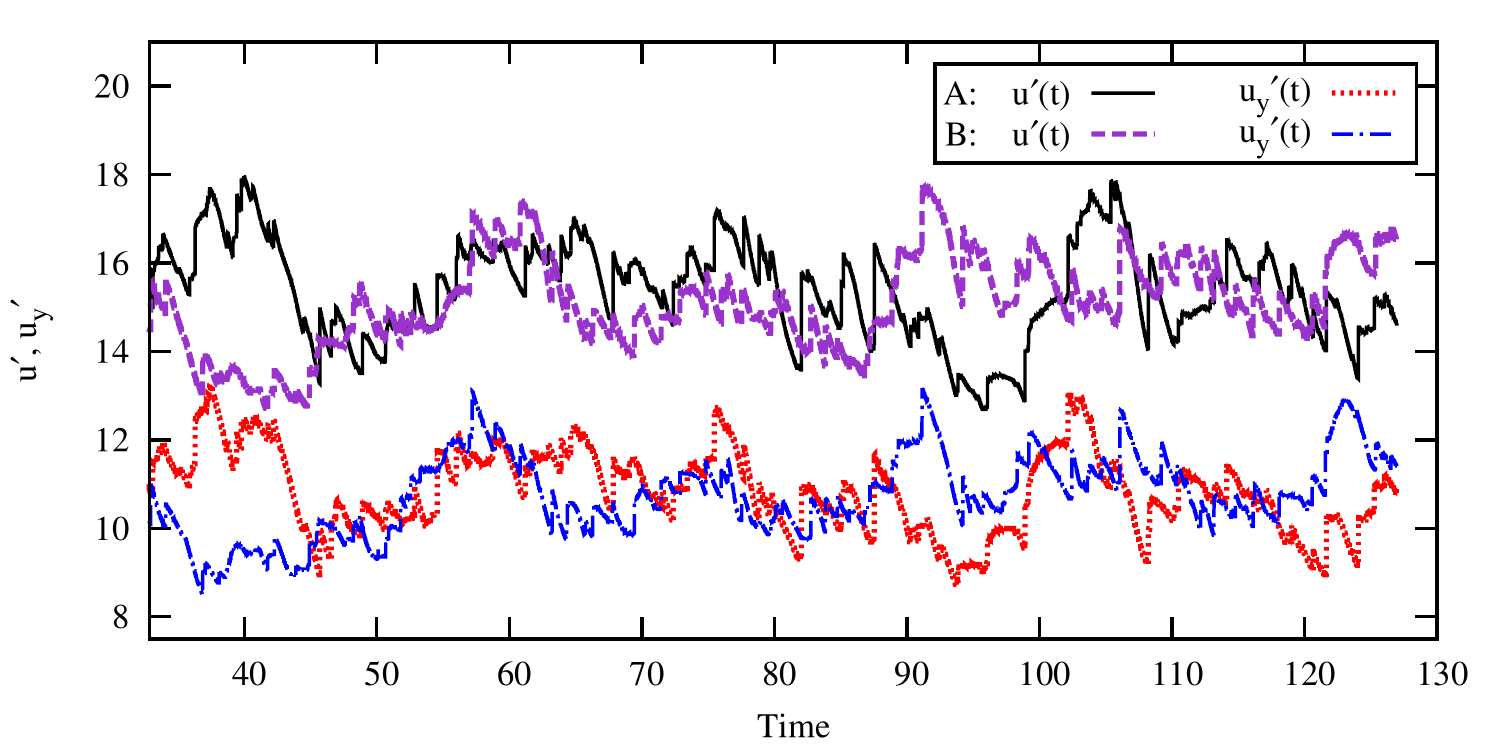}
\end{center}
\caption{Turbulent rms velocities vs. Time.  The turbulent rms velocity, $u'(t)$, measured in the flame brush is shown for Simulation A (solid black line) and Simulation B (dashed purple line).  The turbulent rms velocity in the $y$-direction, $u_y'(t)$, also measured in the flame brush, is shown for A (dotted red line) and B (dot-dashed blue line).  Note that the amplitude of the oscillations is less than the amplitude of the flame speed oscillations.}
\label{fig-4200-s-uprime}
\end{figure*}

In order to compare the data with the predictions of the flame speed models, we measured the turbulent flame speed, $s(t)$, and the turbulent rms velocity, $u'(t)$, for Simulations A and B.  A detailed description of these measurements is given in Section 4 of~\citet{hicks2015}.   To calculate the flame speed, we measured the bulk burning rate~\citep{vladimirova2003a}, 
\begin{equation}
 s(t) = \frac{1}{L^2} \int_0^L \int_0^L \int_{-\infty}^\infty R(T) \,  dy dx dz .
\label{eqn:fspeed}
\end{equation}
which is a measurement of the global consumption of fuel by the flame.  We computed the rms velocity using
\begin{equation}
 u'(t) = \sqrt{ < {u_x(t)}^2 + {u_y(t)}^2 + {u_z(t)}^2 > }
\end{equation}
where $< >$ indicates the spatial average over the volume between the top-most and bottom-most extent of the $T=0.5$ to $T=0.8$ contour range that also satisfies the criterion $T > 0.5$.  So, $u'(t)$ is based on spatial averaging within the ashes.  The rms velocity in the $y$-direction is likewise given by $u_y'(t) = \sqrt{ < {u_y(t)}^2> }$.  Averages are computed over the statistically steady portion of the time series.  Error bars are based on a rolling average procedure which is described in Section 4.3 of~\citet{hicks2015}, and are an estimate of the uncertainty associated with averaging over an oscillating time series.   Both of the new simulations, A and B, were run long enough to take meaningful averages. 

The turbulent flame speeds for the new simulations, A and B, are shown in Figure~\ref{fig-4200vs4210}.  For both simulations, the flame speed varies in a complex, non-periodic way with large oscillations.  These large oscillations are due to competition between the vigorous creation of surface area by the RT instability and the destruction of surface area by burning. The strong RT instability generates the largest average flame speeds in the parameter study with $s_A = 13.86$ and $s_B=13.12$.  Figure~\ref{fig-4200-s-uprime} shows measurements of the turbulent rms velocity ($u'$) and the turbulent rms velocity in the y-direction ($u_y'$).  These velocities also show oscillations, but not as large as those of the flame speed.  The time-averaged rms velocities measured, $u_A'=15.39$ and $u_B'=15.09$, mean that Simulations A and B have the largest time-averaged Reynolds numbers in the parameter study, $Re_A=985$ and $Re_B=966$.   

\subsection{Turbulence-Based Flame Speed Models Comparisons}

\begin{figure*}
\begin{center}
\includegraphics[height=7in,angle=270]{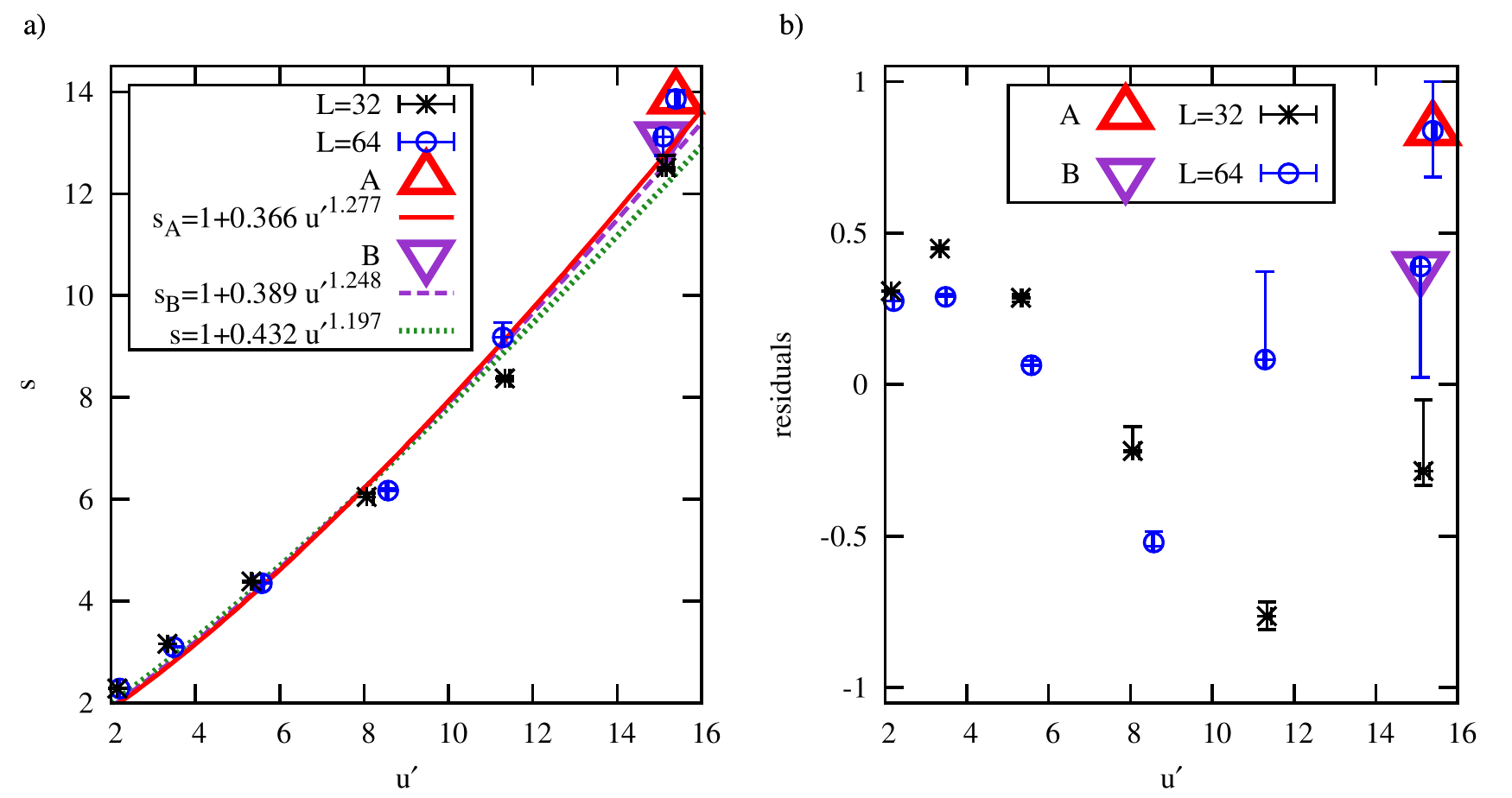}
\end{center}
\caption{Burning Velocity Diagram: Power Law Model ($s=1+C u'^n$).  Here, $s$ is the time-averaged turbulent flame speed and $u'$ is the time-averaged rms velocity in the flame brush.   Each simulation is represented by one data point.  Part (a) shows a comparison between the original least-squares fit of the data without Simulations A or B (dotted green line, $s$) and fits including A (solid red line, $s_A$), or B (dashed purple line, $s_B$).  Simulation A is the red triangle and B is the inverted purple triangle.  Simulations with $L=32$ are represented by black asterisks; simulations with $L=64$ are represented by blue circles.  Adding the new data makes the curve more concave up.  Part (b) shows the residuals of the fit that includes A (red line).  The residuals show a pattern, indicating that, while the power law is useful for demonstrating concavity, it should not be used as a general flame speed model. Note that B is shown on this plot, although it was not included in this fit.}
\label{fig-powerlaw-residuals}
\end{figure*}

\begin{figure}
\begin{center}
\includegraphics[height=3.5in,angle=270]{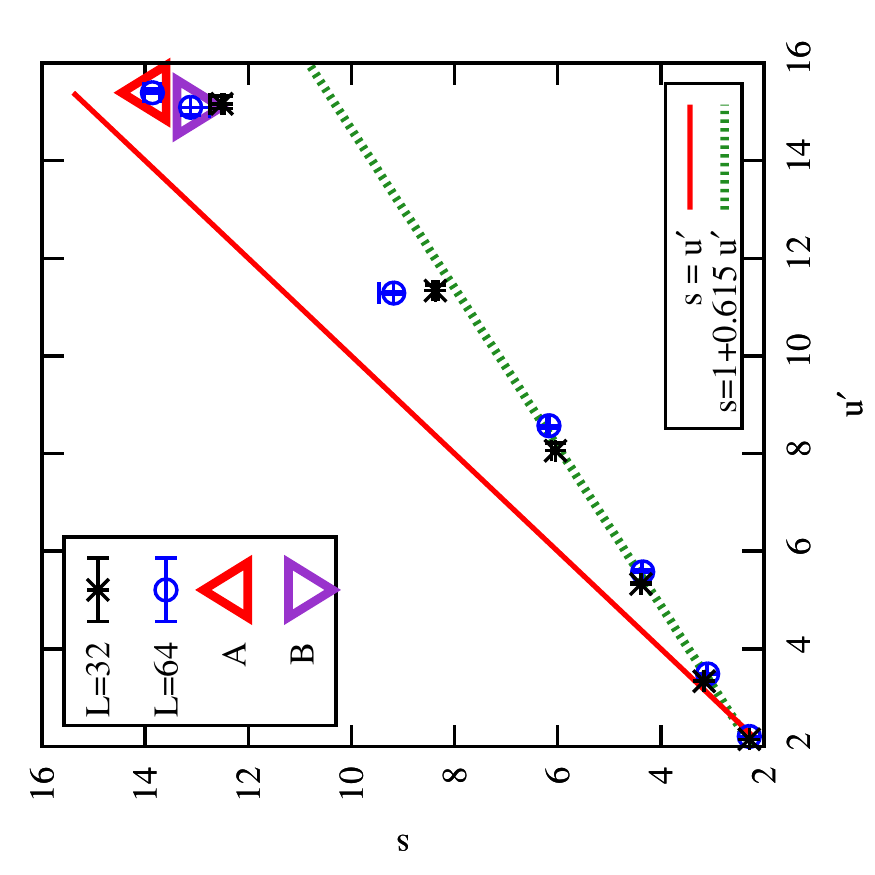}
\end{center}
\caption{Burning Velocity Diagram: Linear Models.  This figure shows two simple linear models: $s=u'$ (solid red line) and $s=1+ C u'$ (dotted green line).  $C$ is determined by a least squares fit of the eight simulations with $u' \lesssim 9$ that show a clearly linear dependence.  The addition of Simulations A (red triangle) and B (inverted purple triangle) do not change our previous assessment (\citet{hicks2015}, Fig. 11)  that neither model fits the data well.}
\label{fig-linearmodels}
\end{figure}

\begin{figure}
\begin{center}
\includegraphics[height=3.5in,angle=270]{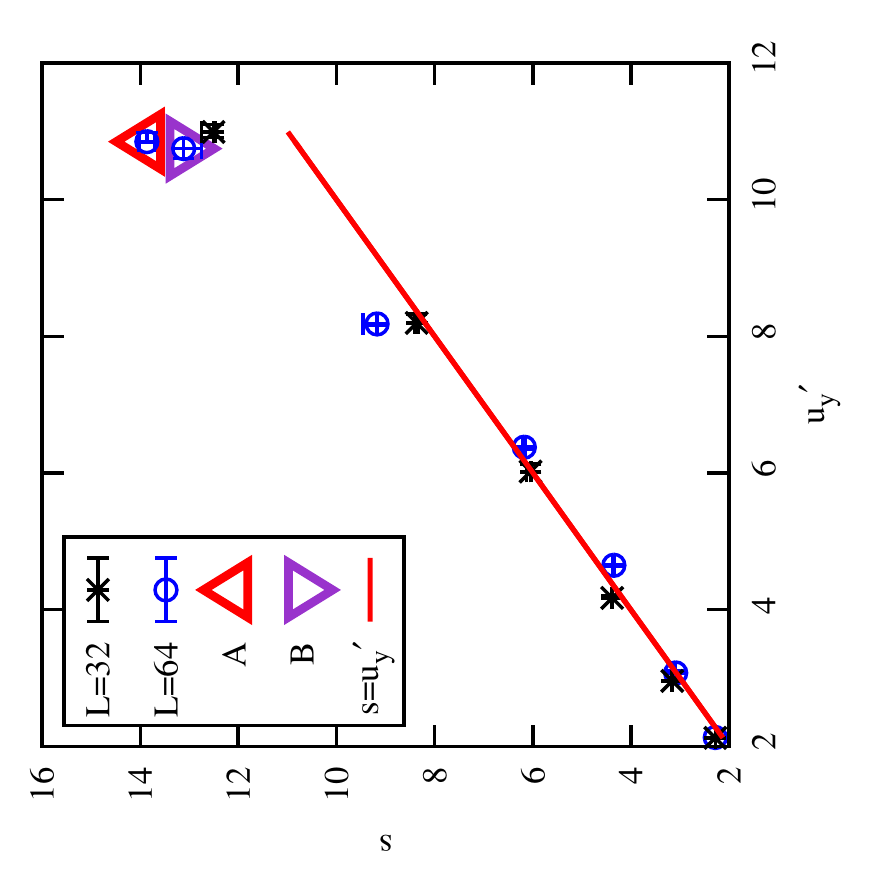}
\end{center}
\caption{Burning Velocity Diagram based on $u_y'$, the time-averaged turbulent rms velocity in direction of flame propagation.  Although the model $s=u_y'$ fits the data well at low $u_y'$ it underestimates the data at high $u_y'$. Simulations A (red triangle) and B (inverted purple triangle) have a much higher flame speed than the model predicts.}
\label{fig-linearmodels-uy}
\end{figure}

\begin{figure}
\begin{center}
\includegraphics[height=3.5in,angle=270]{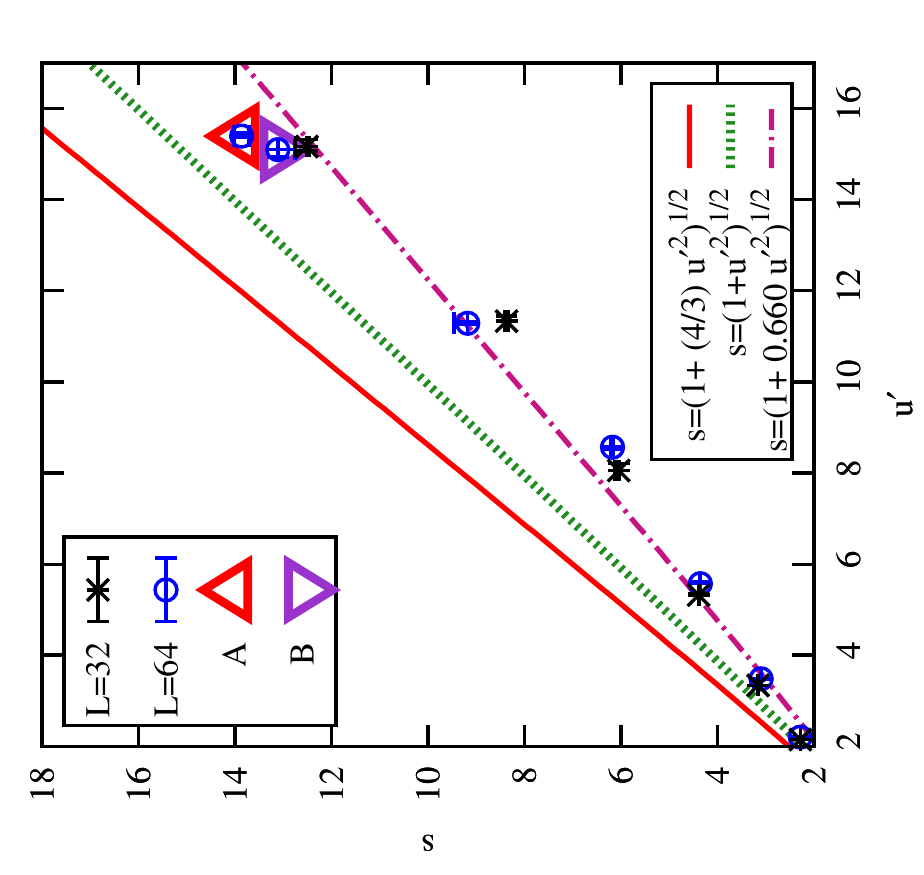}
\end{center}
\caption{Burning Velocity Diagram: Scale Invariant Models.  This figure shows a comparison between the simulation data and the scale invariant model, $s = (1 + C_t {u'}^2)^{1/2}$.  This model with $C_t=4/3$  (solid red line) is commonly used in Type Ia simulations.   $C_t=1$ (green dotted line) was also considered in the formulation of the subgrid model.  Both of these models overestimate the flame speed, even for the new data.  $C_t=0.660$ (dot-dashed purple line) is the best fit including Simulation A, but the pattern of its residuals indicates that it should not be extended to higher $u'$.}
\label{fig-schmidtmodel}
\end{figure}

In this section, we extend our tests from \citet{hicks2015} of turbulence-based subgrid models (Turb-SGS) to include the new $L=64, G=16$ simulations.  After a brief discussion of turbulent flame speed models and their adaptation to the Type Ia problem, we will determine whether the principle finding of \citet{hicks2015} still holds after the new data are added.  Finally, we will discuss specific Turb-SGS models in light of our new results.  

Type Ia Turb-SGS models incorporate findings from the field of turbulent combustion.   Turbulent combustion is the study of flames traveling through turbulence.   One of its main purposes is to find expressions for the turbulent flame speed ($s$) in terms of other system parameters, like the turbulent rms velocity ($u'$).   Originally, it was hoped that a single universal expression for the flame speed in terms of the turbulent velocity, $s(u')$, might be found.   However, it is now clear that such an expression is unlikely to exist, because factors like reaction chain details, system geometry, flame stretch, apparatus geometry and flame quench can influence the turbulent flame speed.  So, different flame speed relations exist for different systems and those relations may involve other parameters in addition to $u'$.   There is no single, coherent theory of turbulent combustion that can make flame speed predictions in novel situations.   

The lack of a unified theory of turbulent combustion makes it difficult to use first principles to select an appropriate turbulent combustion model for Type Ia SN.  Many factors complicate model selection.   Turbulent combustion flame speed models are based on terrestrial flames, which aren't strongly Rayleigh-Taylor unstable. None of the commonly studied classes of turbulent flames match Type Ia supernovae flames geometrically.   Most importantly, turbulent flames travel through upstream homogeneous and isotropic turbulence, while RT unstable flames generate their own more complex turbulence, which is quickly washed downstream from the flame front.   All of these complications mean that prospective Turb-SGS models can't simply be assumed to be correct; they must be tested.

In \citet{hicks2015}, we tested the predicted scalings of three types of Turb-SGS models and showed that they did not match the scaling of our simulated data.   In particular, we considered the shape of both the models and the data on the so-called ``burning-velocity'' diagram, the plot of $s$ versus $u'$.   Practically all theoretical models for turbulent combustion predict either a linear relation between $s$ and $u'$ or a sublinear relation, where the curve on the burning velocity diagram appears to ``bend down'' (see Figure 6 in \citet{hicks2015}).   Surprisingly, we showed in \citet{hicks2015} that our simulated data were \textbf{concave up} instead of concave down.   This makes RT unstable flames unlike every kind of turbulent flame, so we argued that RT unstable flame speed models shouldn't be drawn from turbulent combustion theory.    Of course, our study left open the possibility that RT flames could behave more like turbulent flames at higher $Re$.   The highest $Re$ of any simulation in \citet{hicks2015} was $Re\approx720$.   We were left with the question -- would adding the data from an additional simulation at higher $Re$ make our fit more concave up, or would it start to flatten out and approach a linear model?

After adding the data from Simulation A, we find that the best fit power law model becomes even more concave up than before.  Figure \ref{fig-powerlaw-residuals} shows the whole parameter study, with A marked by a red triangle and B by an inverted purple triangle.  Simulations with $L=32$ are shown with black stars; simulations with $L=64$ are shown with blue circles.   We fit the data to a power law model, $s=1+C u'^n$, with fitted parameters $C$ and $n$.   In \citet{hicks2015}, we found a best fit to the data (which didn't include Simulations A or B) of $s=1+0.432 u'^{1.197}$, which is concave up.  This is shown in the figure as a dotted green line.   After adding A, we now find a new best fit (shown as a solid red line) of $s=1+0.366 u'^{1.277}$.   This is more concave up than our previous fit, which can be seen by taking the ratio of the second derivatives of the two fits, which is $1.271 u'^{0.08}$.   Using the data from Simulation B, instead of A, gives a best fit of $s=1+0.389 u'^{1.248}$  (shown as a dashed purple line) and a second derivatives ratio of $1.182 u'^{0.051}$.    In either case, we see that adding a new, higher $Re$ simulation makes the data at least slightly more concave up than before.   This poses a problem for all of the Turb-SGS models, which predict either linear or concave down behavior and shows how unusual these flames are.  

In \citet{hicks2015}, we compared three basic types of Turb-SGS models to our data, and demonstrated the problems caused by fitting concave-up data with linear or concave-down models.   None of the models fit the data well.  Here, we briefly compare those same models with measurements from our updated parameter set and show that the addition of the $L=64, G=16$ data mostly makes the fits worse. 

The first type of model that we considered was a linear model, which was partially the basis for a Turb-SGS model used by \citet{niemeyer1995}.   This is the simplest of all proposed turbulent flame models, and goes back to the early turbulent combustion studies by \citet{damkohler1940}.  As in \citet{hicks2015}, we compare our data with three linear models:  $s=u'$, $s=1+C u'$ (where C is a fit parameter), and $s=u_y'$.   Figure \ref{fig-linearmodels} shows a comparison between the simulation data and the first two of these linear models on the burning velocity diagram.   $s=u'$ is shown as a solid red line;  this model generally overestimates the flame speed.   However, Simulations A and B are closer to the model line than the $L=32, G=32$ (L32G32) simulation (the black point nearest A and B).  This is one case where the addition of new data makes the model a better fit for the data.  How successful would this model be at even higher $u'$?   

The model $s=1+C u'$ is shown as a dashed green line in Figure \ref{fig-linearmodels}.  The fitting constant, $C$ was found by fitting the eight simulations with $u' \lesssim 9$ that visually show a linear trend, and therefore is not changed by the addition of new simulations.  This model seriously underestimates the flame speeds of the simulations with the highest $u'$.     

Figure \ref{fig-linearmodels-uy} shows a comparison with the model $s=u_y'$.   This model fits the flame speed well at low values of $u_y'$ but severely underestimates the flame speed at high $u_y'$, especially for the new data points.    So, the new simulations don't qualitatively change our previous results; these linear models are still not good fits for the data.

The second type of model that we considered in \citet{hicks2015} was a scale invariant model, which is the basis of the complex LES subgrid model developed and used by \citet{schmidt2005, schmidt2006a, schmidt2006b}.   The model is based on a flame speed relation derived by \citet{pocheau1992,pocheau1994}, $ s = \left[1 + C_t \left( u' \right)^n \right]^{1/n}$, where $n=2$ and $C_t$ is a parameter that is typically set to either $1$ or $4/3$, although, ideally it would be fit for \citep{schmidt2006b}.  A comparison between this model and the data is show in Figure \ref{fig-schmidtmodel}.   The figure shows the model with three different values of $C_t$:  $4/3$ (solid red line), $1$ (dotted green line), and the best fit value including Simulation A, $0.660$ (dot-dashed purple line).  The best fit using Simulation B instead of A is $C_t=0.646$ (not shown).  The data as a whole are still underestimated by both the $C_t=4/3$ and $1$ models.  The new data points are closer to these models, but without more data it is impossible to say whether either of these models could be valid at large values of $u'$.   The best fit line, $C_t=0.660$, fits the data fairly well at low $u'$, but underestimates the value of the new data points.   This is because this model is close to linear at large $u'$, whereas the data are concave up.   It is important to note that the addition of either A or B to the dataset increases the fit constant $C_t$ from the value $C_t=0.614$ found in \citet{hicks2015}.   This, along with the clear pattern in the residuals of the fit, is a sign that this model should not be extended blindly to higher values of $u'$.   Overall, scale-invariant flame speed models are not a good fit for the flame speed measurements.   The values $C_t=1$ and $C_t=4/3$, which have been used in full star simulations, overestimate the flame speed even for our new simulations at higher Reynolds number.

The third type of model that we considered in \citet{hicks2015} was a bending model, which was the basis of the LES turbulence-flame interaction model formulated by  \citet{jackson2014} based on \citet{colin2000,charlette2002, charlette2002a}.   This model was specifically formulated to produce the concave-down bending seen in many terrestrial flames.   The formulation in  \citet{jackson2014} was complex, so in \citet{hicks2015} we compared our simulation data to two other different bending-type models from turbulent combustion theory.   We will not show those comparisons here with the new data points, but only note that the concave-down behavior of all bending models (including  \citet{jackson2014}) makes them a poor fit for our concave-up data.

Overall, the addition of the $L=64, G=16$ simulation data to our parameter study did not change the qualitative conclusions of \citet{hicks2015}.   RT unstable flame data are concave up on the burning velocity diagram, unlike turbulent flame data and models, which are either linear or concave down.   With our new data points, the RT flame trend is even more concave up than we previously found.  This suggests that RT unstable flames are fundamentally different than turbulent flames.   Tests of linear, scale-invariant, and bending models still show that these models do not fit either the shape or the values of the flame simulation data well. As discussed in Section 3, the velocity that would be induced by gas expansion across the flame is most likely too small to affect these results. Overall, the current evidence shows that flame speed models from traditional combustion shouldn't be used for RT unstable flames.

\subsection{RT Flame Speed Model Comparison}

\begin{figure*}
\begin{center}
\includegraphics[height=6in,angle=270]{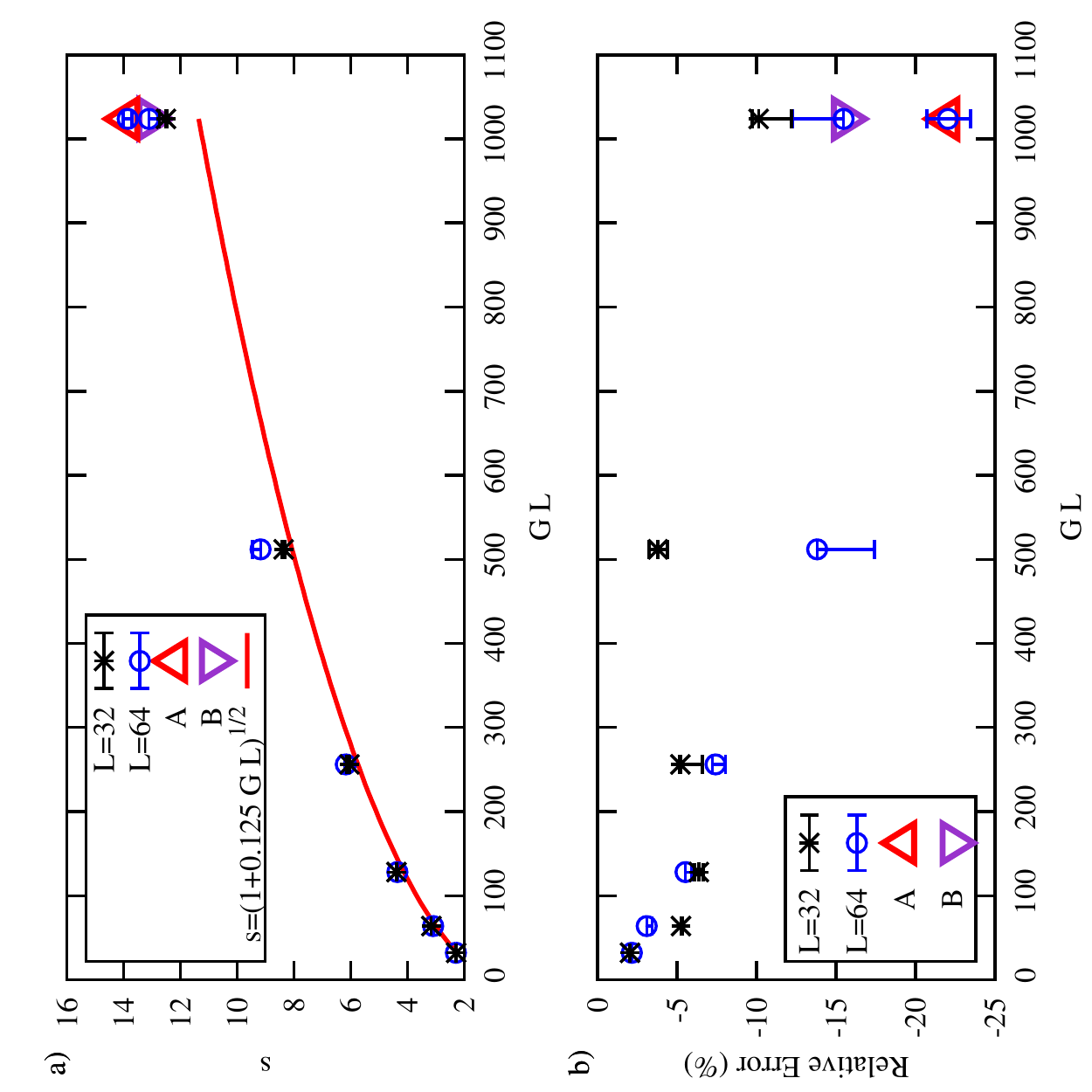}
\end{center}
\caption{Rayleigh-Taylor Flame Speed Model compared with the simulation data.  Part (a) shows a direct comparison between the RT flame speed model prediction (solid red line) and the time-averaged flame speeds measured from the simulations (black asterisks for $L=32$, blue circles for $L=64$).  New simulations A and B are shown as a red triangle and inverted purple triangle respectively.   Part (b) shows the relative error (RE) between the predicted value (PV) and the simulation data (SD), defined as RE=100*(PV-SD)/PV.  The error bars in both plots represent an uncertainty based on averaging over a finite-time oscillating time series (see~\citet{hicks2015}, Section 4.3).}
\label{fig-rt}
\end{figure*}

The RT subgrid (RT-SGS) model assumes that the flame speed is set by the Rayleigh-Taylor instability.   In this model, the flame speed is set by competition between the generation of new flame surface area by the RT instability and destruction of surface area by geometrical effects, like flame collisions~\citep{khokhlov1995, zhang2007}.   The RT instability controls the rate of both processes.   The predicted average flame speed in 3D is $s = s_o \sqrt{1+0.125 G L}$, with the flame speed expected to oscillate around this value as surface area creation and destruction fight for supremacy.   The basic $\sqrt{GL}$ scaling (with a low $G$ correction) was tested in 2D up to GL=512 by \citet{V03,V05} and up to $GL=16,384$ by \citet{hicks2013} and in 3D by  \citet{zhang2007} up to $GL=2786$ and found to model the flame speed well.

However, in \citet{hicks2015}, we showed that the RT scaling matched the flame speed measurements for our simulations at low $GL$, but underestimated the flame speed for two simulations at higher $GL$.    Specifically, we showed the average flame speed of simulation L32G32 ($GL=1024$) was larger than the RT prediction by about $10\%$ and that simulation L64G8 ($GL=512$) was larger than the RT prediction by about $14\%$.  One of our major motivations for carrying out simulations at $L=64, G=16$ was to see whether flames with this parameter set would also travel more quickly than the RT model predicts.   

So, do flames with $L=64, G=16$ travel faster than the RT prediction?    The answer is ``yes'';  the average flame speed of Simulation A is $13.86$ and the average flame speed of B is $13.12$ (see Figure \ref{fig-rt}).   Both of these values are substantially higher ($22\%$ and $15.5\%$, respectively) than the RT predicted value, $s=11.36$.  These measurements confirm our previous conclusions and give us more confidence that we are detecting a real physical effect.  

The magnitude by which the flame speed exceeds the RT prediction likely depends on the flame reaction model.  More peaked reaction rates (like the CF88 reaction) are less susceptible to flame stretching, which decreases the local burning rate.  Generally, flame stretching acts counter to effects like cusp burning and flame collisions that increase the local burning rate and that could be responsible for the excess flame speeds measured here.  This means that RT unstable CF88 flames could potentially travel even faster than RT unstable bistable flames.

\section{Conclusions}

This study was motivated by the surprising findings of~\citet{hicks2015}, which used a large parameter study to show that RT unstable flames are different from turbulent flames in several ways.  First, RT unstable flames are thin rather than thick when turbulence is strong.  This is contrary to the regime predictions of turbulent combustion theory.  Second, the flame speed is concave up rather than concave down on the burning velocity diagram, which is different from turbulent combustion models (which are either linear or concave down).  Together, these two findings suggested that Turb-SGS flame speed models, which are based on traditional turbulent combustion theory, are not physically appropriate for RT unstable flames.  In addition, we found in~\citet{hicks2015} that the RT flame speed model works well when the RT instability is weak but underpredicts the flame speed when instability is strong.  We concluded that full star simulations of Type Ia supernovae should generally use the RT-SGS model but that a new model was likely needed for flames in the outer part of the star.  However, these findings depended on simulations for which $Re\lesssim720$; we did not explore higher Reynolds numbers.

In this paper, we presented two new fully resolved simulations (A and B) at higher Reynolds number to test the conclusions of~\citet{hicks2015}.  Both simulations were at $L=64$, $G=16$ and they had average Reynolds numbers of $Re_A=985$ and $Re_B=966$, respectively.  The strength of the RT instability for these simulations was the same as for the $L=32$, $G=32$ simulation from~\citet{hicks2015} because they have the same $GL=1024$, but the new simulations had larger $Re$ because the box size was larger.  The larger box size also meant that A and B had a smaller $Ka$ than the L32G32 simulation, but it was still the second largest $Ka$ in the parameter study.  The flame speed of both simulations oscillated wildly in time because of the competition between the creation of surface area by the strong RT instability and the destruction of surface area by burning (see Figure~\ref{fig-4200vs4210}).

Analysis of the two new simulations confirmed and extended the basic findings of~\citet{hicks2015}.  The average flame width between $T=0.1$ and $T=0.9$ for both new simulations is about $3.5$, which is less than $\delta^{0.9}_{0.1} = 4.394$.  These flames are stretched thin by the RT instability instead of thickened by turbulence.  There was no sign of any transition to the thin reaction zones regime; these flames remain in the flamelets regime although $Ka > 1$.  The addition of the new flame speed data to the parameter study made the flame speed curve on the burning velocity diagram more concave up.  This meant that the linear, scale invariant and bending flame speed models adapted from turbulent combustion theory fit the data qualitatively less well because they are either linear or concave down.  Adding the new data makes certain models (the linear model, $s=u'$, and the scale-invariant model, $s = \left[1 + C_t \left( u' \right)^2 \right]^{1/2}$ with $C_t=4/3,1$) quantitatively better, but all three of these models still overestimate the flame speed for the new simulations.  In particular the $C_t=4/3$ model overestimates the flame speed for the new simulations by about 30 percent.  The best fit for $C_t$ changed from $0.614$~\citep{hicks2015} to $0.660$ using Simulation A or $0.646$ using Simulation B.  The pattern of residuals for these fits shows that even this ``best fit'' model is not a good overall fit for the data.  Generally, the scalings of the Turb-SGS models do not work well for this data set.  In addition, the RT model substantially underestimates the flame speeds for the new simulations.  This is the third set of parameters for which we have seen this effect, which increases our confidence that the RT flame speed model generally fails when the RT instability is strong.  Overall, these results extended the basic trends seen in~\citet{hicks2015}; the flames in the new simulations were thin, the flame speed on the burning velocity diagram became more concave up, and the RT model underpredicted the flame speed.

Our new results support the general physical picture of RT unstable flames developed in~\citet{hicks2015}.  The dynamics of the flame seem to be mostly controlled by the RT instability, instead of by self-generated turbulence.  Several pieces of evidence suggest this.  First, the flame speed for the new simulations is dominated by large scale oscillations, which suggests that RT self-regulation is occurring.  These oscillations are much larger and on a longer timescale than those of $u'$.  Second, the flame front visually appears to be dominated by RT bubbles and mushroom fingers (see Figure \ref{fig-flamefront}).  Third, the flame is stretched thin by the RT instability instead of thickened by turbulence.  Finally, turbulent flame speed models generally fail to predict the flame speed well.  The fact that the RT instability seems to control the flame dynamics doesn't mean that the turbulent flow field has no effect on the flame.  It probably does act with the RT instability to shape the flame front.  However, turbulence seems to be washed downstream fast enough that it doesn't affect the actual flame structure.  In addition, it is clear that the RT instability sets the strength of the turbulence.  The turbulent rms velocity, $u'$, depends only on the strength of the RT instability, which scales with $GL$.  Simulations with the same $GL$ have similar values of $u'$ even if their flame speeds are different.  For example, the values of $u_A'=15.39$ and $u_B'=15.09$ for the two new simulations are close to $u'=15.16$ for the L32G32 simulation, although their average flame speeds span a much wider range, from $s_A=13.86$ and $s_B=13.12$ to $s =12.51$ for L32G32.  So, the RT instability seems to control both the flame dynamics and the energy budget of the system.

The conclusion that the RT instability controls the flame dynamics probably does not depend on our use of the Boussinesq approximation and the bistable model reaction.  The gas expansion driven Landau-Darrieus instability has been shown in other work~\citep{ropke2003, bell2004a, ropke2004a, ropke2004b} to be uncompetitive with the RT instability, and the maximum velocities induced by gas expansion ($\approx s_o$) are much smaller than the turbulent velocities for these simulations.  In addition, there is no sign that the overall dominance of the RT instability depends on the reaction rate.  While the highly peaked CF88 reaction is less likely to be thinned by the RT instability, it is also less susceptible to disturbance by turbulent eddies.  The competition of these effects should be explored in future work.

The RT instability is clearly not the only factor setting the flame speed, because the RT flame speed model fails when the RT instability is strong (see Figure~\ref{fig-rt}).  In~\citet{hicks2015} (Section 4.6) we hypothesized that the RT model underestimates the flame speed because it doesn't account for enhanced burning in cusps.  Cusps are regions of high curvature on the flame surface that can be formed by normal-direction propagation of the flame surface (the Huygens mechanism,~\citet{zeldovich1966}), by turbulence~\citep{khokhlov1995,poludnenko2011a}, or by the RT instability~\citep{hicks2015}.  Cusps increase the local consumption of fuel because they geometrically focus thermal flux.  Colliding flame sheets can also focus thermal flux.  In practice, cusps and dense flame packing go together.  If there are enough cusps on the flame surface, or if the flame is densely packed, the global flame speed will be noticeably higher.  In~\citet{hicks2015}, we indirectly evaluated this hypothesis by checking that the order of magnitude of the flame speed enhancement seemed reasonable, and by comparing two simulations with the same GL, but different flame speeds.  Visual inspection of both new simulations shows both cusps and dense flame packing, especially when the flame speed is high (see Figure~\ref{fig-flamefront}), so this hypothesis still seems reasonable.   However, a much more rigorous analysis is needed to truly test it and to quantify the relative contributions of cusps and dense flame packing. In particular, it is important to determine how these relative contributions depend on the reaction rate.

The implications of these results for the choice of Type Ia and Iax subgrid model are basically the same as stated in~\citet{hicks2015} and apply when the convective turbulence is not strong.  The RT-SGS model should be a good choice until either the RT instability or turbulence acts at the scale of the flame width.  As this isn't expected until the flame reaches the outer parts of the white dwarf, the RT-SGS model should be appropriate for most of the deflagration phase.   In the later stages of evolution, a new subgrid model is likely needed.  Turb-SGS models are probably only appropriate if the turbulence encountered by (not generated by) the flame is strong enough to completely overwhelm the RT instability.  In addition, the commonly used scale invariant model formulated by \citet{schmidt2005,schmidt2006a,schmidt2006b} is likely to be setting the flame speed too high.  Finally, because our flame width results make detonation by the Zel'dovich gradient mechanism following a transition to the thin reaction zones regime less likely, the focus on alternative detonation scenarios should continue.

There is a huge amount of work still to be done on RT unstable flames and subgrid models for Type Ia and Iax supernovae.  It would doubtless be informative to expand this parameter study to larger $L$ and $G$. At higher $L$ we could explore the regime where turbulence is strong but $Ka < 1$; at higher $G$ we could search for the transition to thin reaction zones when $Ka >> 1$ and explore whether there is some upper bound to the amount of ``extra'' flame speed possible and to the amplitude of the flame speed oscillations.  White dwarfs have very small $Pr$, so simulations should explore $Pr<1$.  The simulations in this parameter study could be repeated with C-O flames at low Mach number or with full compressibility to see how more realism would complicate our simple picture.   Finally, simulations of RT unstable flames propagating through pre-existing turbulence of various strengths could be used to measure the flame speed for ignition scenarios where convective turbulence in the white dwarf is strong.  

There are also many interesting investigations that can be undertaken with the current parameter study.   In particular, we plan to study the variability of the flame properties with time and the details of self-regulation.  Does it make sense to use the average flame speed for subgrid models when the PDF of possible flame speeds is so wide?  As a follow-up to this work, we are currently using local measurements of the flame (like curvature) to explore its local structure and determine whether RT-generated cusps or flame packing can explain the increased global flame speed.  Connecting local and global properties is the next step in understanding RT unstable flames.

\section*{Acknowledgments}
Thank you to R. Rosner for originally introducing me to Rayleigh-Taylor unstable flames and to N. Vladimirova for introducing me to the \textsc{Nek5000} code and providing setups and scripts when I first started working on this problem.  I also thank N. Vladimirova and R. Rosner for interesting discussions that have influenced my thinking over the years.  I am very grateful to P. Fischer and A. Obabko for making \textsc{Nek5000} available and for giving advice on using it.  Thank you to F.X. Timmes for making his EOS routine freely available online.  Thank you to S. Tarzia for proofreading and content suggestions. I also thank the anonymous referee for insightful comments and questions.  This research used resources of the Argonne Leadership Computing Facility, which is a DOE Office of Science User Facility supported under Contract DE-AC02-06CH11357.   Specifically, Mira was used for Simulations A and B and Cooley was used to visualize them.  This work also used the Extreme Science and Engineering Discovery Environment (XSEDE), which is supported by National Science Foundation grant number ACI-1548562 \citep{xsede}.  Continued storage of the entire parameter study, including Simulations A and B, on Ranch at the Texas Advanced Computing Center (TACC) is made possible through XSEDE allocation TG-PHY170024.   Additional analysis was performed on Comet at the San Diego Supercomputer Center (SDSC) also with XSEDE allocation TG-PHY170024.  Visualizations (Figure 1) in this paper were generated with \textsc{VisIt} \citep{visit}.  \textsc{VisIt} is supported by the Department of Energy with funding from the Advanced Simulation and Computing Program and the Scientific Discovery through Advanced Computing Program. Gnuplot was used to generate Figures 2-11.

\bigskip

\noindent This is a pre-copyedited, author-produced version of an article accepted for publication in MNRAS following peer review. The version of record: E P Hicks, Rayleigh-Taylor unstable flames at higher Reynolds number, Monthly Notices of the Royal Astronomical Society, Volume 489, Issue 1, October 2019, Pages 36-51, is available online at: https://doi.org/10.1093/mnras/stz2080.  



\bibliographystyle{mnras}
\bibliography{bibliography}





\bsp	
\label{lastpage}
\end{document}